\documentstyle[eqsecnum,aps,amstex,amssymb,epsfig,float]{revtex}

\begin{document}              
\draft
\title{Comparison of filters for detecting gravitational wave bursts in 
interferometric detectors}

\author{Nicolas Arnaud, Matteo Barsuglia, Marie-Anne Bizouard, Violette Brisson,
Fabien Cavalier, Michel Davier, Patrice Hello\footnote{email : hello@@lal.in2p3.fr}, 
Stephane Kreckelbergh, Edward K. Porter and Thierry Pradier}

\address{ Laboratoire de l'Acc\'el\'erateur Lin\'eaire, B.P. 34,B\^atiment 200,
Campus d'Orsay,
91898 Orsay Cedex (France)\protect\\}

\maketitle

%================================================================
%================================================================

\begin{abstract}

Filters developed in order to detect short bursts
of gravitational waves in interferometric detector outputs are compared according to three main points. 
Conventional Receiver Operating Characteristics (ROC) are first built for all the considered filters 
and for three typical burst signals. Optimized ROC are shown for a simple pulse signal in order
to estimate the best detection efficiency of the filters in the ideal case, while realistic ones
obtained with filters working with several ``templates'' show how detection efficiencies
can be degraded in a practical implementation.
Secondly, estimations of biases and statistical
errors on the reconstruction of the time of arrival of pulse-like  signals are then given for each filter.
Such results are crucial for future coincidence studies between Gravitational Wave
detectors but also with neutrino or optical detectors.
As most of the filters require a pre-whitening of the detector noise, the sensitivity to a non perfect
noise whitening procedure is finally analysed. 
For this purpose  lines of various frequencies and amplitudes are added to a Gaussian white noise
and the outputs of the filters are studied in order to monitor the excess of false alarms induced
by the lines.
 The comparison of the performances of the different filters finally show that
they are complementary rather than competitive.

\end{abstract}


%================================================================
%================================================================

\baselineskip = 2\baselineskip

%================================================================
%================================================================
\section{ Introduction}
%================================================================
%================================================================
Long baseline interferometric detectors of gravitational waves (GW) 
\cite{ligo,virgo,geo,tama}
are currently taking their first data. The preparation for data analysis of  compact 
binary 
inspiral signals, the most promising source of GW to date, with 
these new instruments has been in progress for a long time , as well as for periodic 
sources 
(see eg \cite{schutz} for a review). 
The effort concerning the search for burst sources is more recent.
The expected GW burst sources are primarily massive star collapse
either with neutron star (NS)
\cite{monchmeyer,bona1,zwerger,rampp,relat,relat2,relat3} or black hole (BH) formation
\cite{stark}. The duration of such events is at most a few milliseconds
and the simulated amplitudes do not exceed a few $10^{-23}$ (NS formation)
or $10^{-22}$ (BH formation) for sources located at about 10 Mpc.
With these typical amplitudes, interferometric detectors of the first generation
have no hope of ``seeing'' such events if they occur beyond the Galaxy 
\cite{relat3,arnaud}.
Other sources of GW bursts are NS binary  \cite{ooh,rasio,ruf,janka,jr,jr2}
or BH binary mergers \cite{bbh1,bbh2,bbh3,bbh4,bbh5}, for which a large amount of 
effort 
is currently underway in order to predict plausible waveforms.
More exotic, but detectable by first generation interferometers, are possible
GW bursts emitted by cosmic strings \cite{cusps}.

All the predicted sources of GW bursts are in fact characterised by a very
rough knowledge of the emitted waveforms. Unless simulations of core collapse
or binary mergers can provide accurate waveforms (which seems currently doubtful), 
the use of matched
filtering, which would be the optimal method in case of perfect knowledge
of the waveforms, is not possible. Sub-optimal filtering methods are then required
for detecting GW bursts. Such methods have been developed in differents groups
in the last few years. The Excess Power Monitoring has been built upon in successive
versions \cite{flana,powermonit,power2,vicere}.  Time frequency methods
are also planned for burst signal or noise non-stationarity 
detection \cite{bala,mohanty,fabbroni,sylvestre}.
In our group, we have developed
filtering methods with the idea of being as robust as possible with respect to the 
possible signal waveforms \cite{arnaud,moriond,pra,prados}.
We have also introduced a benchmark test in order to compare different filters \cite{arnaud}
in a given situation. This benchmark is however incomplete if we want to fully understand
the different methods. The goal of this paper is then to compare the different proposed filters.
A filter based on moving average\cite{papou} is added to the
bank of previously published filters. The definitions of the
filtering methods are first recalled. 
We intentionally discard some of them, such as Bin Counting \cite{arnaud}
or the Norm of the Autocorrelation \cite{pradthese}, which are clearly not competitive
compared to the others. The
efficiency curves (detection probability vs false alarm rate)
of the different filters for generic burst signals with various signal-to-noise 
ratios (SNR) are
then computed. 
We first use optimal versions of the filter (in the sense that the filter parameters 
are matched
to the signal)  and secondly realistic implementations for each
filter, with different ``templates'' working in parallel.
In the following part, the timing properties of the filters
(bias and statistical timing accuracies) are studied with respect to both
signal SNR and signal width. The sensitivity of the filters 
to the noise whitening quality is finally discussed.

%=============================================================
%=============================================================
\section{Filters for detecting GW bursts}
%=============================================================
%=============================================================

In this section, we firstly describe the noise model we will use in all the following
and then we enumerate and briefly describe the filtering methods.

%=========================================================================
\subsection{The noise model}
\label{noisedef}
%=========================================================================
Throughout the paper, we assume that the noise is Gaussian and white with zero 
mean. The standard deviation of the noise is then :
\begin{equation}
\sigma = \sqrt{ S_h f_0 \over 2 },
\label{sigma}
\end{equation}
where $f_0$ is the sampling frequency and $S_h$ is the one-sided spectral 
density of the noise.
For numerical examples, we take $f_0=20$ kHz (Virgo sampling rate)
and $\sqrt{S_h} \simeq 4\times 10^{-23}$ $/\sqrt{\mathrm{Hz}}$, which is about 
the minimum value of the foreseen noise spectral density of the Virgo 
interferometer \cite{virgosens}; this choice is correct since the minimum is located
 in the frequency 
range for the expected burst sources of GW. The fact that we choose
 Gaussian noise is not essential, but simply convenient for the design 
of the filters. Deviation from gaussianity 
will produce for example an excess in the rate of false alarms and it will then
be possible for example
 to retune the algorithms thresholds according to the real noise statistics. 
In the frequency
range of interest, above a few 100 Hz,
 the Virgo noise sensitivity curve is rather flat, although 
not exactly white. Most of the filtering methods presented here and in \cite{arnaud} 
require a whitening of the noise \cite{cuo,cuo2}, which is foreseen for the
Virgo data processing output. In the following, we normalise the 
noise level by its standard deviation, so that we are dealing with a Gaussian noise 
with zero mean and unit standard deviation. 
We also denote the data $x_i$ at sample times $i/f_0$.
Let's recall also that we conventionally define a Signal to Noise Ratio (SNR) after 
filtering
as (filter output -$m_o$) / $\sigma_o$ where $m_o$ and $\sigma_o$ are the mean and 
standard
deviation of the filter output in the absence of a signal. Of course a SNR can have 
different
dimentionality depending on whether the filter is linear or quadratic for instance.
This shows at least that the SNR is an ambiguous criterion
when we are interested in comparing different filters, linear or not.

%=========================================================================
\subsection{The Norm Filter}
%=========================================================================

The Norm Filter \cite{arnaud} is a simple version of the Excess Power statistics
\cite{flana}. It is based upon a monitoring of the local signal energy in a moving 
window
\begin{equation}
y_k = \sum_{i=k}^{k+N-1} x_{i}^2.
\end{equation}
Under this form, the filter appears to be non linear with a single parameter, the moving
window size $N$.
In presence of noise only, $y_k$ is distributed as a $\chi^2$ variable with N
degrees of freedom (mean $N$ and standard deviation $\sqrt{2N}$).
The variable 
\begin{equation}
y_k^{NF} = \sqrt{2y_k}-\sqrt{2N-1} = \sqrt{2\times \sum_{i=k+1}^{k+N} x_{i}^2}- \sqrt{2N-1}
\end{equation}
can be very well approximated by a standard normal variable, if $N\succsim 30$ \cite{spiegel}.
This is the definition of the Norm Filter (NF). We note of course that $y_k^{NF}$ is normalised
so that it is well a SNR.
The response of the NF to a test signal is displayed in Figure 1. The NF is able to recover
about 70\% of the optimal SNR in this example.

%=========================================================================
\subsection{The Slope filters and ALF}
%=========================================================================

A family of filters based upon fitting a straight line to the data has
been proposed \cite{prados}.
The two (non independant) results of the fit, namely the line slope $a$ and
the offset value $b$
\begin{align}
a &=  {<tx>-<t><x> \over <t^2>-<t>^2}, \\
b &=  <x>-a<t> = <x> - {<tx>-<t><x> \over <t^2>-<t>^2} <t>
\end{align}
where $<x>= {1\over N} \sum_{i=1}^N x_i$ and $t_i = i/f_0$, can be used as linear filters
with variances
\begin{align}
\sigma_{a}^2 &=  {12 f_0^2 \over N(N^2-1)}, \label{sa} \\
\sigma_{b}^2 &=  {4N+2 \over N(N-1)}.
\end{align}

The normalized Slope and Offset Filters, namely 
$y_k^{SF}=a/\sigma_a$
and $y_k^{OF}=b/\sigma_b$, can be uncorrelated
by diagonalisation
of the covariance matrix, yielding
\begin{equation}
y_k^\pm = { y_k^{SF} \pm y_k^{OF} \over \sqrt{2(1\pm\alpha)} },
\end{equation}
with correlation coefficient
\begin{equation}
\alpha =  -\sqrt{{ 3\over2} \left({N + 1\over 2N + 1 }\right)}.
\end{equation}
The two uncorrelated filters can finally be combined and the ALF (non linear) filter
is obtained :
\begin{equation}
y_k^{ALF}=(y_k^+)^2+(y_k^-)^2 = {(y_k^{SF})^2+(y_k^{OF})^2-2\alpha y_k^{SF}y_k^{OF} \over 1-\alpha^2}. \end{equation}
In presence of noise only, ALF is well approximated with
a $\chi^2$ distributed random variable with 2 degrees of freedom, hence a mean and 
a standard deviation both equal to 2.
The only parameter is $N$ in all cases. Again, simple recurrence
relations can be used in the successive calculations of the SF and OF (and so for ALF) outputs.
The responses of SF, OF and ALF to a test signal are shown on Figure 1.
The moving window size for each filter is optimally chosen to be $N=140$, that is about 7
times the signal half-width as stated in \cite{prados}.
We see that OF is performs better in detecting the signal than SF in this example.
As already noticed, ALF (quadratic filter) and the other (linear) filters don't have the
same ``dimension'', since ALF outputs are proportional to the energy of the signal and not to 
the amplitude, hence the much larger SNR for ALF.

%=========================================================================
\subsection{The Mean Filter }
%=========================================================================

We now introduce the Mean Filter (MF), which is nothing but the filtering by moving average
or boxcar average \cite{papou}.
The Mean Filter computes the mean of the data in a moving window
\begin{equation}
y_k^{MF} = \frac{1}{N}\sum_{i=k}^{k+N-1} x_{i}.
\end{equation}
It is a linear filter with a single parameter, the window size $N$. In presence
of noise only, $y_k^{MF}$ is distributed as a Gaussian random variable with
zero mean and standard deviation $1/\sqrt{N}$. The computation of the filter outputs
is very fast, as trivial recursive relations between $y_k^{MF}$ and $y_{k+1}^{MF}$
can be used. So the moving window can be allowed to move bin by bin, without concern
for CPU time.
The response of the MF to a test signal is displayed in Figure 1. We see that the MF is able
to recover almost all of the optimal SNR in this example.

A first means of comparing the filters is to benchmark them within common conditions, i.e. subjecting them
to the same GW signals for identical noise conditions.
We first used a benchmark  based on a catalogue of 
78 supernovae signals, simulated by Zwerger and M\"uller (ZM)
\cite{zwerger,webSN} in the axisymmetric case (see \cite{arnaud}and \cite{prados} for details).

The results of this first benchmark for the different filters are collected in Table 1. We recall
those already shown in Ref.\cite{prados} and  add the new one for the Mean Filter. We note that MF 
has a performance similar to the ALF's one.

~

\begin{center}
\begin{tabular}{|c|c|c|c|c|c|c|c|}
  \hline Filter  & Optimal & NF & SF & OF & ALF & {\bf MF }\\
\hline Average distance (kpc) & 27.4 & 11.5  & 11.3 & 15.2  & 22.5& {\bf 20.0} \\ 
\hline Performance           & 1    & 0.46  & 0.49 & 0.59 & 0.81& {\bf 0.78} \\\hline
\end{tabular}
\end{center}
{Table 1 : Distances of detection and performances of the different filters 
in detecting a sample of supernovae signals.
NF = Norm Filter, MF = Mean Filter, SF = Slope Filter, OF = Offset Filter.
The results are extracted from Ref.\cite{prados} except for the MF new one.}

%=========================================================================

%=========================================================================
\section{Efficiency of the filters}
%=========================================================================

Unfortunately the previous benchmark gives only a partial view of filters performances, as
it is computed for a specified false alarm rate. Of course one would wish to extend the comparison
of filters to other false alarm rates, especially for those lying in the likely range
allowed during science runs of interferometers. Such a tool is standard in signal processing,
the so-called Receiver Operating Characteristics (ROC), which displays the curves 
detection efficiency vs the false alarm rate. In the next section, we will compute the ROC
for typical (albeit of course arbitrary) burst signals. This will complete our understanding
of the detection power of the filters.

\subsection{Methodology}

For each filter, we compute the ROC for three distinct typical
burst-like signals. The first
is a Gaussian pulse of half width  $w=1 $ms of the form
\begin{equation}
s(t) = A \,\exp\left( - \frac{(t-t_0)^2}{2 w^2} \right).
\label{gauss}
\end{equation}
The second is a damped sinusoid of frequency $f=1$ kHz and damping time $\tau=1$ ms of
the form
\begin{equation}
s(t) = A \, H(t-t_0) \, 
\exp\left( - (t-t_0)/\tau \right)\, \sin\left(2\pi f (t-t_0)\right),
\end{equation}
where $H(x)$ is the Heaviside step function ($H(x)=0$ if $x<0$ and $H(x)=1$ if $x>0$).

The last waveform is a 
supernova signal from the ZM catalogue (number 6 in order of decreasing simulated
signal energy) \cite{zwerger}. The three signals are displayed in Figure 2.
The amplitude $A$ of each of these three signals is calibrated according
to the corresponding optimal SNR $\rho_0$ (if one of the signals were detected
by optimal filtering with the same noise conditions then the mean optimal SNR would be
$\rho_0$). We have used for the Monte Carlo simulations a data window
of size 2048. For each simulated data window, we first pass the filters with noise only. 
If one of the filters is triggered then we increment its the false alarm counter, 
else we add one of the signals to the noise and look if the filter detects
the signal, in which case its detection counter is incremented.
The efficiency of a filter in detecting one of the signals is then the ratio of the 
number of detections by the number
of noise realisations without false alarm. Meanwhile, the false alarm rate is the ratio
of noise realisations with a false alarm to the total number of noise realisations
and then divided by the data window size, resulting in a false alarm rate per bin.
The data window size (2048 bins) has in fact been chosen large enough to contain
the signals but not too large in order to have a very low probability of having
more than one false alarm in a single window. This obviously may occur only for very high
false alarm rates (not reached in practice).
We first study the case of optimized filters (only one ``template'' matched to the signal
we consider) in order to have information about the maximal efficiencies the filters
can reach. We then study the case of realistic implementations, with several
``templates'' working in parallel. It is worth noting that the ``event'' notion introduced
in \cite{prados} is here automatically taken into account, as the detection algorithm stops
(and the detection counter is incremented)
as soon as the filter output is above threshold. The method is thus independant of the real number
of bins above threshold and of the details of event clustering, as defined in \cite{prados}.

\subsection{ROC for optimized filters}

We first consider optimized filters, in the sense that their
parameters (essentially the moving window size $N$) are optimally matched to the signal.
We consider for this purpose the Gaussian pulse signal (Eq.\ref{gauss}).
With $w =1$ ms, the filters are matched with window sizes $N=40$ (MF and NF)
or $N=140$ (SF, OF and ALF). The ROC for various optimal SNRs are shown
in Figures 3, 4 and 5 ($\rho_0=5$, 7.5 and 10 respectively).
For each SNR value, MF and ALF show very close efficiencies. While much simpler, MF
is able to compete with ALF  for detecting pulse-like signals. 
However this is the ideal situation, all filters being 
optimally implemented with respect to the signal. We will see how this is modified
with realistic implementations in the next section. Concerning ALF, it is again
clear that we gain in combining OF and SF (in ALF), as OF and SF are always less efficient
ALF whatever the false alarm rate or the signal strength. Between OF and SF,
OF is always significantly more efficient than SF. Finally NF appears as non competitive
for detecting short pulses even in its optimized version. In particular, 
in the case of low SNR ($\rho_0=5$) the NF efficiency
is close to zero for practical false alarm rates in the interferometers (say $ < 10^{-6}$).
For higher SNR ($\rho_0=10$), NF can not reach 50\% efficiency in the false alarm rate
range we study. On the contrary for such a SNR, MF and ALF have efficiencies near 100\%
over all the range of false alarm rates.

\subsection{ROC for practical implementations of the filters}

In the previous section, the filters were matched for a single pulse
width. In the real world however, the signal (if any) width will not be known in advance. 
Moreover the signal itself will surely not be a perfect Gaussian pulse. That is why
the filters must be implemented with different ``templates'' in order to conveniently cover
the signal parameter space. For all the filters here, the only parameter
is the moving window size. Thus in practice the filters will be implemented (for
instance in the online trigger system)
with different moving windows in parallel. Such an implementation is shown in Table 2,
where the typical burst width ranges from 0.5 ms to 10 ms. For NF and MF the window
sizes correspond to the signal widths $\Delta t$ ($N=\Delta t \times f_0$), 
while for ALF, SF and OF they correspond to about
3.5 times the signal widths \cite{prados}.

\begin{center}
\begin{tabular}{|c|c|c|c|c|c|c|c|c|c|c|}
\hline signal size (ms)& 0.5 &  0.75 & 1 & 1.25 & 1.5 & 2  & 2.5 & 3.5 & 7.5 & 10 \\ 
\hline MF, NF & 10 & 15& 20 & 25& 30 & 40  & 50 & 70 & 150 & 200 \\ 
\hline SF, OF and ALF & 35 & 50 & 75 & 90 & 105 & 140 & 175 & 250 & 500 &750\\\hline
\end{tabular}
\end{center}
{Table 2 : Choice of the 10 window sizes to be implemented in parallel for MF and
NF, and for SF, OF and ALF. The corresponding typical signal widths are also given.}

~

The ROC for the Gaussian pulse of half-width 1 ms are shown in Figure 6.
The optimal SNR is 5 (the most interesting to exhibit filters differences with the
false alarm rates considered). Note that the signal width corresponds exactly to one
of the window sizes of the implementation ($N=40$ for NF and MF and $N=140$
for the slope filters familly). When compared to Figure 3, the realistic situation
differs from the ideal one in particular for MF. ALF is
the most efficient, followed by OF and SF, while NF is always the worst . The case of MF
is interesting : its efficiency is excellent if matched to the pulse width but
it dramatically decreases if implemented with several ``templates''. The matched
template ($N=40$ here) is always as efficient in detecting the pulse but the other
templates contribute to increase the number of false alarms. For the other
filters the increase of false alarms due to the mismatched templates is much less.
This can be also exhibited by comparing the false alarm rates $\kappa_{50}$ for which
the filter efficiency reaches 50\%. Note that $\kappa_{50}$ can be also
a good quality criterion for comparing the different filters.
The results are shown in Table 3.

\begin{center}
\begin{tabular}{|c|c|c|c|c|c|}
\hline Filter  & ALF &  OF & SF & MF & NF  \\ 
\hline Optimized filters &$10^{-7}$&$2\times10^{-7}$&$2\times10^{-6}$&$10^{-7}$&$>2\times10^{-4}$ \\ 
\hline Realistic filters &$2\times10^{-7}$&$3\times10^{-7}$&$4\times10^{-6}$ &$4\times10^{-5}$ & $4\times10^{-4}$ \\\hline
\end{tabular}
\end{center}
{Table 3 : False alarms rates $\kappa_{50}$
for which a filter efficiency reaches about 50\%
for a Gaussian pulse of half-width 1 ms and optimal SNR $\rho_0 =5$.
The figures are extracted from Figures 3 (optimized filters) and 6 (realistic filters).}

~

The sensitivity of the overall false alarm rate to the number of templates is then
a problem for MF while the other filters seem much more robust with respect to
this aspect. Indeed their efficiency curves shift only slightly to the right
if we compare Figures 6 and 3, or their $\kappa_{50}$ changes by only
a factor roughly about 2. For MF the shift is again very large and its $\kappa_{50}$
changes by a factor of about 400.

The next ROC for the damped sine signal are plotted on Figure 7. The situation is completely
different than before. 
If compared to Figure 6, MF has about the same efficiency, while NF is significantly better.
But for this signal, ALF and parent filters are the worst.
The 50\% efficiencies are obtained for false alarm rates 
$\kappa_{50} \simeq 4\times10^{-5}$ for NF
(about the same as for the Gaussian signal), $\kappa_{50} \simeq 2\times10^{-4}$ 
for NF (twice as good) 
and about $\kappa_{50} \simeq 3\times10^{-4}$ for ALF, that is $10^3$ 
worse than for the Gaussian signal.
In this configuration ALF is not competitive while MF is the best filter in the list.

Finally the ROC for the supernova signal are shown on Figure 8. We find roughly 
the hierarchy first obtained with the Gaussian pulse. ALF and related filters
are the most efficient, MF arrives next and the least efficient is NF.
The 50\% efficiencies are here obtained for false alarm rates about
$\kappa_{50} \simeq 3\times10^{-5}$ for ALF, OF and SF, $\kappa_{50} \simeq 7\times10^{-5}$
for MF and $\kappa_{50} \simeq 5\times10^{-4}$ for NF.

\subsection{Discussion}

The first point to mention is that the relative efficiencies of the different filters
depend strongly on the type of waveform.
For instance ALF is not suited to the detection of the damped sine signal above. 
MF and NF have very roughly the same detection efficiencies
whatever the signal, so in this sense these filters are robust. However ALF can be much more
efficient than MF and NF. In all cases we find 
that ALF is always more efficient than SF and OF. We effectively gain a lot
in combining SF and OF in ALF.
In its optimized version ($N$ matched to the signal size), MF can be as efficient as ALF
(see Figure 3), but the efficiency
falls dramatically with a practical implementation. MF is however the most efficient for the damped
sine signal. NF is never the most efficient for any signal but it appears more efficient than ALF
in detecting the damped sine signal.
All these results are summarized in the Table 4 where the false alarm rates for
50\% detection efficiency for
all the signals are given.

\begin{center}
\begin{tabular}{|c|c|c|c|}
\hline   & ALF & MF & NF  \\ 
\hline Gauss (ideal) & $10^{-7}$ ($\star$)               & $10^{-7}$   ($\star$) & $>2\times10^{-4}$ \\ 
\hline Gauss (realistic)&$2\times10^{-7}$  ($\star$)     & $4\times10^{-5}$ & $4\times10^{-4}$ \\
\hline Damped sine (realistic)&$3\times10^{-4}$ & $4\times10^{-5}$ ($\star$)  & $2\times10^{-4}$ \\
\hline ZM (realistic)&$3\times10^{-5}$  ($\star$)          & $7\times10^{-5}$ & $5\times10^{-4}$ \\
\hline
\end{tabular}
\end{center}
{Table 4 : False alarms rates $\kappa_{50}$
for which the filter efficiency reaches about 50\%
for the 3 signals with  optimal SNR $\rho_0 =5$. For each signal the best $\kappa_{50}$ 
value is marked ( $\star$).
The results for OF and SF are not reported in the Table since their detection efficiencies are always less than ALF ones (by construction).}

~

It may appear surprising that MF peforms better than NF or ALF in detecting the damped sinusoidal waveform, as the
mean of a sinusoid is 0. In fact MF is efficient because one of its ``templates''width is well
adapted to detect one of the signal peaks.
In fact, a cut-off frequency may be associated with each of the templates of length $N$ 
through $f_c=f_0/N$, where $f_0$
is again the sampling frequency. For the implementation shown in Table 2, the largest cut-off frequency
(associated to the lowest value of $N$) is $f_c^{MAX} \simeq 1$ kHz. Below this frequency, 
there will be always at least
one template short enough to pick out only one signal peak. On the contrary, above $f_c^{MAX}$, all the templates
average the signal cycles and the MF output dramatically decreases. A similar behaviour is found
for ALF, due to the fact that the mean slope of a sinusoid with many cycles is zero.
This can be seen in Figure 9, where
the detection efficiencies of MF, NF and ALF for the same false alarm rate are shown as a function of the
frequency of a damped sine signal of damping time $\tau = 100$ ms (long enough to have many cycles when the frequency
is high enough).
We see clearly that ALF and MF are very efficient at low frequency, while they dramatically lose efficiency
as the signal frequency increases. On the contrary the NF efficiency is almost constant, whatever the signal
frequency. This shows some robustness for the NF. We note also that for the signal frequency $f=1$ kHz, MF
is again more efficient than NF and ALF, as in Figure 7, while its efficiency tends towards 0 as the signal
frequency becomes larger and larger.

%=========================================================================
\section{Timing issues}
%=========================================================================

\subsection{Methodology}
The filter timing properties are very important since timing accuracy is necessary
(1) for validating coincidences between GW detectors, and (2)  for reconstructing the signal.
For instance, the signal time delay between VIRGO and LIGO-Hanford is up to 27 ms. Thus, the
 estimation accuracy
of the time of arrival of a signal obtained with a given filter must be much less than this delay.
This timing accuracy is also crucial in the case of coincidences with neutrino detectors.
In the latter case, GW timing accuracy needs to be below 1 ms in order to not
limit the measurement of neutrino masses\cite{neutrinos}.

The definition of a time-of-arrival estimator can be in general non trivial since it can
depends both on the filter and on the signal waveform. That is why we will use in fact a simple 
waveform in order to evaluate the optimal performance of a filter to  measure a signal time of arrival,
keeping in mind that for a real signal the timing accuracy should be degraded.

In order to evaluate the timing accuracy of the filters we proceed as follows. We consider 
as a burst signal
a Gaussian pulse (Eq.\ref{gauss}) with a variable amplitude $A$ (always calibrated according
to the optimal SNR $\rho_0$) and a time width $w$. The signal is buried
in white Gaussian noise and the filters try to detect it. We define the expected time of arrival
as the maximum of the pulse, that occurs at $t_0$ for the signal of Eq.(\ref{gauss}). The first arrival time estimator
considered for the filters is given by the SNR maximum.
A different estimator will be in fact used for ALF.
After many noise realisations we can thus estimate the systematic bias $\Delta t$ (mean of the
distribution of measured times of arrival) if any and the statistical error $\sigma_t$ (RMS
of the distribution) on $t_0$ for a given set of parameters ($w$ and $\rho_0$).

In the case of optimal filtering (here correlation of the noisy Gaussian pulse with an identical
Gaussian pulse template that is nothing but the Peak Correlator \cite{arnaud} with a single template) no systematic bias is found and the statistical error is 
\cite{networknous}:
\begin{equation}
\sigma_t \simeq 0.145\; {\mathrm ms} 
\left( \frac{w}{\mathrm 1\; ms}\right)\left(\frac{\rho_0}{10} \right)^{-1}
\end{equation}

We note that the statistical error is linear with respect to both $w$ and $1/\rho_0$.
This timing accuracy is in fact the best that can be achieved (optimal filtering). 
We see that for the canonical example, $w=1$ ms and $\rho_0=10$, $\sigma_t \simeq 0.15$ ms,
well below the time delay between interferometers or the required accuracy for coincidences
with neutrinos. The question is then to investigate 
if the studied suboptimal filters retain acceptable
timing accuracies. The bias and statistical errors on the measure of $t_0$ a priori depend
on the amplitude $\rho_0$ and on the width $w$ of the pulse signal. For every filter, we will then
study both $\Delta t$ and $\sigma_t$ as functions of $\rho_0$ and $w$. We will use the matched
versions of the filters (window size $N$ matched to the signal width $w$) so the numbers
given are to be considered as upper limits, that are
the best achievable in principle with the filters.

\subsection{Norm Filter}

For the Norm Filter, we first find a systematic bias
$\Delta t = \frac{N}{2} \times$ sampling time. This shift is simply related to the window
size $N$ and can be easily corrected for. The correction can itself be incorporated in the filter
definition so it plays no role. The statistical error has less trivial relations with
the signal parameters. Figures 10 and 11 (left panel) show the behaviour of $\sigma_t$
as function of $\rho_0$ and $w$ respectively. The statistical error does not behave linearly
either with $1/\rho_0$ or with $w$, contrarily to the optimal filter. In log-log scales however,
the curves are linear and the slopes can be obtained from a least squares fit. 
They differ substantially from unity and are of course larger than the (plus or minus unity) slopes
found in the case of optimal filter.
The results can be combined into a single expression for the statistical error
\begin{equation}
\sigma_t \simeq 0.273\; {\mathrm ms} 
\left( \frac{w}{\mathrm 1\; ms}\right)^{1.07}\left(\frac{\rho_0}{10} \right)^{-0.72}.
\end{equation}

\subsection{Mean Filter}

For the Mean Filter, we find exactely the same systematic effect as for NF, 
$\Delta t = \frac{N}{2} \times$ sampling time. Again this bias can be corrected for 
and is unimportant.
The statiscal error $\sigma_t$ is first found to behave linearly with respect to the signal
width $w$ (so better than NF). But, as for NF, $\sigma_t$ is not linear with respect to $1/\rho_0$, as shown
in Figure 11 (mid panel). The slope of the curve $\sigma_t$ vs $\rho_0$ in log-log scales is about $-0.68$,
well above $-1$ as for the optimal filter, but a little worse than NF.
 We can combine the results into a single expression:

\begin{equation}
\sigma_t \simeq 0.246\; {\mathrm ms} 
\left( \frac{w}{\mathrm 1\; ms}\right)\left(\frac{\rho_0}{10} \right)^{-0.68}.
\end{equation}

We note that for the canonical values $\rho_0 =10$ ans $w=1$ ms, we obtain
$\sigma_t \simeq 0.25$ ms, about the same as for NF (about 0.27 ms), which is of course larger
than the optimal filter statistical error, but not much larger.

\subsection{ALF and related filters}

For SF and ALF, the response to a pulse signal shows two peaks as seen in Figure 1.
The peak maxima are in principle symmetric relatively to $t_0$. We consequently
define the time of arrival new estimator as $\tilde{t}_0 = (t_1+t_2)/2$ where $t_1$ and $t_2$ are the
time locations of the two peaks. For OF the situation is 'normal' (a single
maximum) as for NF and MF and the time of arrival estimator is not modified.
For the 3 filters SF, OF and ALF we find again the usual systematic
$\Delta t = \frac{N}{2} \times$ sampling time, that we can correct for.
Then the behaviour of the statistical error $\sigma_t$ is found to be linear with respect
to the signal size $w$, even for the non linear filter ALF.
Finally, as for NF and MF, the three filters do not have a linear behaviour with respect to
$1/\rho_0$. The Figure 11 (right panel) shows the curves $\sigma_t$ vs $\rho_0$ for
ALF (results for SF and OF are not displayed). In each case the slope 
in log-log scales is well below 
$-1$ and similar
to what has been found for NF and MF, that is around $-0.7$.
The results are combined into the following formulas

\begin{equation}
\sigma_t^{SF} \simeq 0.244\; {\mathrm ms} 
\left( \frac{w}{\mathrm 1\; ms}\right)\left(\frac{\rho_0}{10} \right)^{-0.69},
\end{equation}

\begin{equation}
\sigma_t^{OF} \simeq 0.311\; {\mathrm ms} 
\left( \frac{w}{\mathrm 1\; ms}\right)\left(\frac{\rho_0}{10} \right)^{-0.68},
\end{equation}

\begin{equation}
\sigma_t^{ALF} \simeq 0.253\; {\mathrm ms} 
\left( \frac{w}{\mathrm 1\; ms}\right)\left(\frac{\rho_0}{10} \right)^{-0.71}.
\end{equation}

\subsection{Discussion}

The first and important point is that all the suboptimal filters studied in this paper
have good timing accuracies. The systematic bias for the time of arrival are trivial
and can be easily corrected for. The statistical errors are of course larger than in the case of
optimal filtering but are still acceptable. For the canonical example, $w=1$ ms and $\rho_0=10$
(see Table 5),
the statistical error is about 0.15 ms for optimal filtering, about the same for ALF, SF, MF and NF
(around 0.25 ms) and the worst is obtained for OF, about 0.3 ms that is twice the optimal value.
In any case, in this example, the timing accuracy is well below 1 ms for all the filters.
\begin{center}
\begin{tabular}{|c|c|c|c|c|c|c|}
\hline Filter  &Optimal& ALF &  OF & SF & MF & NF  \\ 
\hline $\sigma_t$ (ms) &0.15 & 0.25& 0.31& 0.24 &0.25 & 0.27\\\hline
\end{tabular}
\end{center}
{Table 5 : Statistical errors for the time of arrival estimation for a Gaussian pulse
with width $w=1$ ms and amplitude normalised to $\rho_0=10$.}

~

We then note that all linear filters behave linearly with respect to the signal width. For
the non linear filters, ALF also displays such a behaviour (as a descendant of two linear filters)
while NF does not. All the suboptimal filters have a non-linear behaviour with respect
to $1/\rho_0$, the best one here being NF (slope closest to -1), but the difference
between the filters is not really significant as the slopes range from -0.68 for MF and OF
to -0.71 for ALF and -0.72 for NF.

Finally, balancing all the aspects, the best filter concerning timing accuracy seems to be ALF,
NF being penalized by its non-linearity with respect to the signal width.
This conclusion was not a priori obvious, considering the definition and the 
broad response of the ALF
filter (see Figure 1). However this optimistic conclusion should be moderated. Indeed,
 this result has been obtained with a simple waveform, a single peak.
With a more structured signal (a supernova signal), the situation is in fact less favourable.
The first problem is to properly define a time of arrival estimator. Different estimators have been tested in a Monte Carlo
simulation for determining the time of arrival of ZM signals with ALF \cite{pradthese} : 
time of maximum SNR, time of first bin above threshold,
average time between two SNR peaks etc... All of them have been found to be biased. 
Moreover the bias may strongly depend on the waveform
type (type I, II or III as classified by Zwerger and M\"uller \cite{zwerger}). The smallest bias, averaged 
over all the waveforms of the ZM catalogue, is about 0.5 ms for signal SNR of 5. Type
III signals have the largest contribution to this bias (with biases around 1.6 ms in average).
This shows that the ALF timing accuracy (bias + statistic error) can be in some cases
significantly larger than 1 ms.
This can have serious consequences, for example, in the case of coincidences with neutrino detectors. Indeed,
if the GW timing accuracy is worse than about  1 ms, the determination of neutrino masses is
degraded \cite{neutrinos}.

%=========================================================================
\section{Effect of lines on the filter false-alarm rates}
%=========================================================================

The suboptimal filters studied in this paper 
require a pre-whitening of the noise. In reality,  noise whitening is never perfect
and we need to quantify the filters sensitivity to imperfect whitening. This has been already
be done for ALF \cite{prados} but needs to be extended to the other filters. For this purpose,
about 100 hours of {\bf Gaussian} white noise data are simulated and a single frequency component
of the form $A \sin (2\pi f t)$
(that can mimic a large line residual) is added. 
We then filter the data (Gaussian white noise+line)
with the different algorithms (including ALF in order to be able to compare the filters
in exactly the same situations) using slicing windows
and record the effective number of false alarms for each one as functions of $A$ and $f$.
 In the following,
a nominal false alarm rate of $10^{-6}$ is chosen. So for 100 hours of data sampled
at 20 kHz about $7200\pm 85$ false alarms are expected (the standard deviation of about 85
is estimated from a binomial law). For instance, we find $n_0=7138$ false alarms with the MF
for the 100 hours of only Gaussian white noise. For each filter, we then increase the sine amplitude
$A$, measure the new number of false alarms $n$ each time and compute the relative excess
of false alarms with respect to the ideal situation ($A=0$), i.e. the quantity $(n-n_0)/n_0$.
The results for NF, MF and ALF are given in Figures 12, 13 and 14 respectively.
We investigate as examples four different frequencies, 0.6 Hz corresponding to the
natural pendulum mode frequency of the suspended mirrors in VIRGO, 100 Hz, 200 Hz
and 400 Hz corresponding to power line harmonics or wire resonances.
The filters window sizes have been chosen to be $N=50$ for both NF and MF and
$N=170$ for ALF, so that all window sizes correspond to the same matched signal
size (about 2.5 ms).
The results for the three filters are quite different. 
For NF the curves for the false alarms excess vs $A$ are about the same whatever the line frequency.
In contrast, for MF and ALF, the false alarm excess strongly depends on the line frequency,
from very large excess at low frequency to lower or even vanishing at higher frequencies.
For MF we find again the effect of the cut-off frequency associated to the
window size $N$, $f_c = f_0 / N$. 
Above this cut-off frequency, the line does not increase anymore the number of false alarms
and has simply no effect on MF performance, since,  by construction, MF
averages the fast oscillations in the window and the net effect is zero.
The same phenomenon appears also, but to a lesser extent, for ALF.
We can now set a specification for the whitening procedure such as the excess in
false alarm rate does not exceed the nominal value more than 10\% .
For NF this implies that the whitened line amplitudes must be lower than 7-8\% of the ideal Gaussian
white noise RMS $\sigma$ for all frequencies. For MF and ALF, the effect depends on the frequency.
For example, for the important 0.6 Hz frequency in VIRGO, the specification is that the line amplitude
should be less than about 2\% of the noise RMS for MF and about 1\% for ALF. For increasing
frequencies the specifications become less and less stringent until they can be totally
relaxed above the cut-off frequency.
Such specifications (from 1\% to 10\% of ideal noise RMS for line amplitudes) seem quite severe.
Let's define the noise flatness \cite{cuo}
\begin{equation}
\xi = \frac{\exp\left(\frac{2}{f_0} \int_0^{f_0/2} \ln S_h(f) df \right)}{
\frac{2}{f_0} \int_0^{f_0/2} S_h(f) df},
\end{equation}
where $S_h(f)$ is the noise one-sided power spectral density (PSD). The flatness is such that $\xi \in [0;1]$, the extremal
values being reached for a very peaky PSD ($\xi \simeq 0$) and for a white PSD ($\xi \simeq 1$).
 The most stringent requirement
(for the 0.6 Hz pendulum mode) corresponds to a maximal line amplitude at about 1\%
of the background Gaussian and white noise. If we convert this in terms of noise flatness
we obtain a specification \cite{pradthese} that reads $\xi \gtrsim 0.97$. 
This may seem very demanding at first sight, but it is  within the reach of existing whitening algorithms
that can already achieve a level of $\xi > 0.98$ \cite{cuo}.

Asking for no more than 10\% extra false alarms due to imperfect noise whitening
is in fact not very demanding as far as detection probability is concerned. Indeed, we can correct
for the false alarm excess of 10\% (in order to recover the desired false alarm rate) by fine tuning
of the detection thresholds.
But relaxing the requirements to 20\% or 50\%  does not really change the situation at least for
the low frequency lines (see Figures 15-16). For MF and ALF the specifications on low frequency line amplitudes
will always be {\it a few }  \% of the background noise RMS since the increase of false alarms is a very sharp
function of the line amplitude.

Of course the number of false alarms considered so far is the raw number, i.e. the number
of outputs above threshold. However as the consecutive outputs of filters are not independant and once a filter
is triggered by noise alone, we find in practice not a single  but a number
of consecutive outputs above threshold and corresponding to the same false alarm event. We can thus redefine
an {\it event} as proposed in \cite{prados}. A false alarm event is then no longer
a simple filter output above theshold but a cluster of successive outputs above threshold, at most separated
by the correlation length of the filter. For example, for 
the simulated 100 hours of data, without lines added, we find for ALF (with threshold
=27.85)
7208 raw false alarms that are resolved in 1170 false alarm events, so a reduction factor about 6. When lines are added
it has been checked that one order of magnitude on the false alarm rate is gained if we take the new definition
of a false alarm event. This can then relax somewhat the constraints on the line amplitudes.
As an example, for the 0.6 Hz line with amplitude $A=7.5\times10^{-2} \times \sigma$, we find more than 63,000
false alarms in the 100 hours of noisy data, giving only 6361 false alarm events, still below the nominal
false alarms number (7208) for the 100 hours of data free of lines.

To be complete let's finally mention that the shapes of the curves {\it relative false alarm event excess}
vs {\it line amplitude} are in fact very close to the previous curves (with the brut false alarm excess).
That means that, if we first have set the desired absolute false alarm rate to some value and if we set
the specifications on the whitening quality in asking again no more than 10\% false alarms in excess,
we will find about the same specifications for remaining line amplitudes, whatever the false alarm
definition.

%=========================================================================
%=========================================================================
\section{Conclusions}

In this paper, the principles of GW burst detection algorithms previously studied in our group have
been recalled. The Mean filter, based on moving average, has been introduced.  Their performances for detecting a set of simulated
supernova signals have been completed and summarized. This was until now the only way to compare the different filters.
We have then considered three other tools or criteria in order to fully understand and compare the filters.
Firstly we have systematically made use of the conventional ROC in detecting typical burst signals. Secondly,  
 the timing
properties (systematic biases and statistical errors in timing reconstruction) have been evaluated.
 Finally we have studied
the effect of a non perfect whitening (remaining spectral lines) on the effective false alarm rate of the filters.

Concerning the old benchmark,  ALF and the Mean Filter 
show the best performance in detecting the supernova signals of the Zwerger and M\"uller catalogue.
When looking at the ROC, ALF and the Mean Filter are still ahead provided they can be optimized, meaning that 
the signal length is known in advance. In practical situations, where banks of filters are used in order
to cover the signal sizes space, the situation is not as clear. In particular, for
 short damped sine signals, ALF is much less efficient than the Mean Filter and even the Norm Filter.
A characteristic feature of the Mean Filter as well as the Norm Filter is their relative robustness : their
efficiency curves are very similar from one signal to another. ALF is much less robust; it is 
in general (much) better but in some cases it is the worst performing.
Concerning the timing issues, all the filters have similar timing accuracies, worse (but no more than a factor  2)
than the optimal filter timing accuracy.
The main concern is maybe about the whitening quality which is required by most of the algorithms.
Indeed the remaining line amplitudes are required to stay below a fraction of the ideal background Gaussian white
RMS. Fortunately the redefinition of a false alarm event allows one to gain almost one order of magnitude
on the effective resulting false alarm rate.
Finally, taking everything into account, it is difficult to state that one filter is better than the others.
Rather than establishing some hierarchy, as done for example with our previous benchmark, the important
conclusion of this paper is that the different filters are in fact complementary. This is indeed manifest
when we again look at the different ROC : a single filter can not be the most efficient for all burst signals.
Then rather than using a single 'preferred' filter, it is advisable to operate a battery of different filters
having their own qualities and defects. The next step is then to develop a strategy to find the best
use of all the filters in the battery.

As we have presented in this paper an unbiased way to estimate the performance of filters, 
we would like to suggest its application in the context of development of GW burst detection methods. 
This could be very valuable in order to
directly compare  the performances of different filters and their complementarity or redundancy. 

%=========================================================================
%=========================================================================
\baselineskip = 0.5\baselineskip  
%=========================================================================

%=========================================================================
 
%============================================================================
\newpage
%============================================================================
%---------------------------------FIG 1--------------------------------------
\begin{figure}

\centerline{\epsfig{file=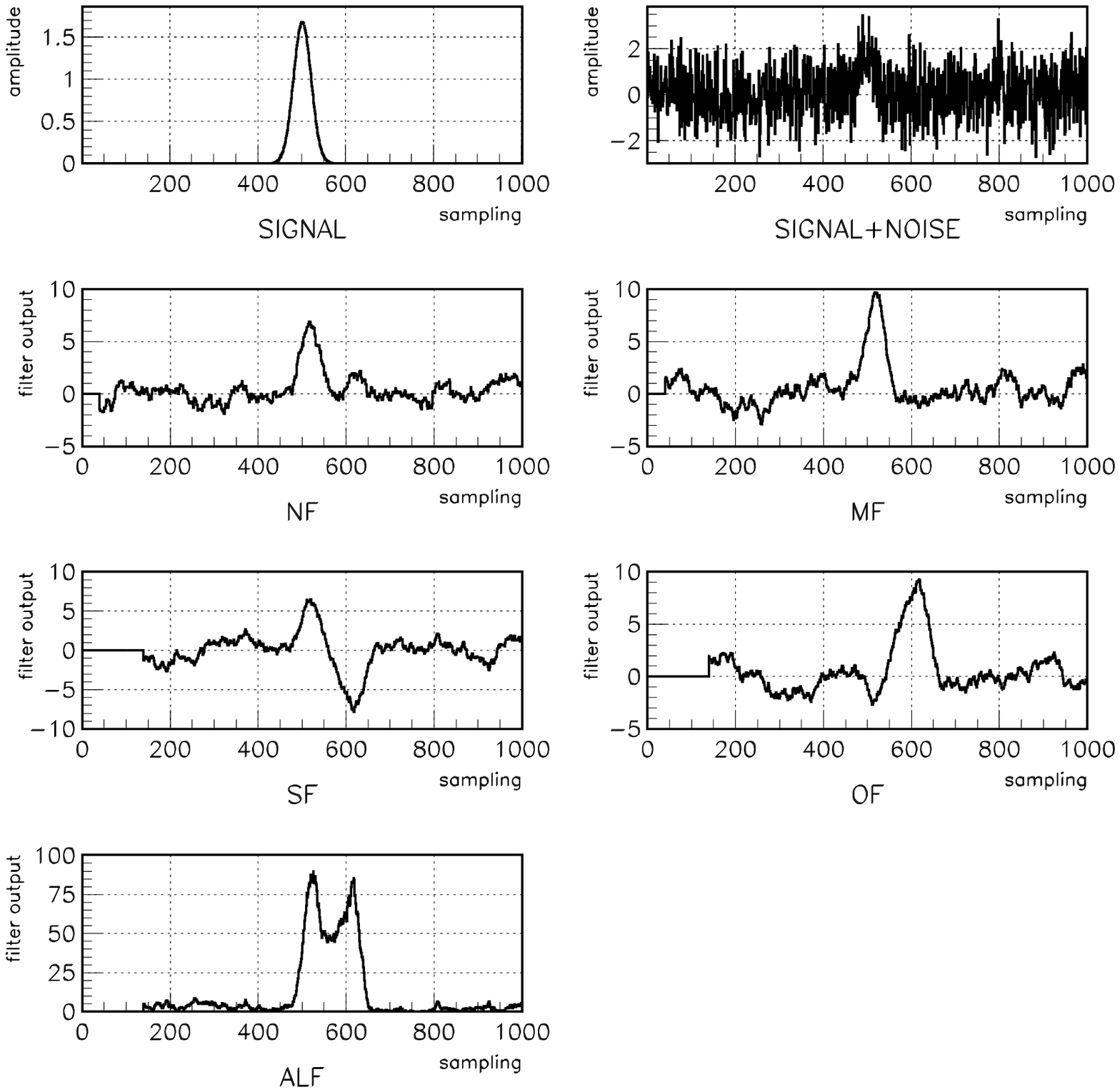,height=15cm}} 
\caption{Responses of the filters to a Gaussian burst signal of 
half width 1 ms (upper left panel) embedded in noise (upper right panel)
which is assumed to be white Gaussian with zero mean and unity standard deviation.
The 5 following plots show the responses of the Norm Filter (NF), the Mean Filter (MF),
the Slope Filter (SF), the Offset Filter (SF) and ALF.
For each filter the moving window size is chosen to be optimal :
$N=40$ (corresponding to the signal width $N/f_0 \simeq 2$ ms) for NF and MF
and $N=140$ for ALF and related filters.
In this example, the optimal SNR is 10;
the maximal NF SNR is about 6.9, the maximal MF SNR is about 9.7, the maximal SF SNR is about 7.8
and the maximal OF SNR is about 9.3. The maximal ALF (quadratic) SNR is here about 40.
We note in each case the obvious time delay between the signal peak and the output maximum.
This will cause a trivial bias when we study the filters' time resolution (see section IV).}

\end{figure}
%----------------------------------------------------------------------------
%---------------------------------FIG 2--------------------------------------
\begin{figure}
\centerline{\epsfig{file=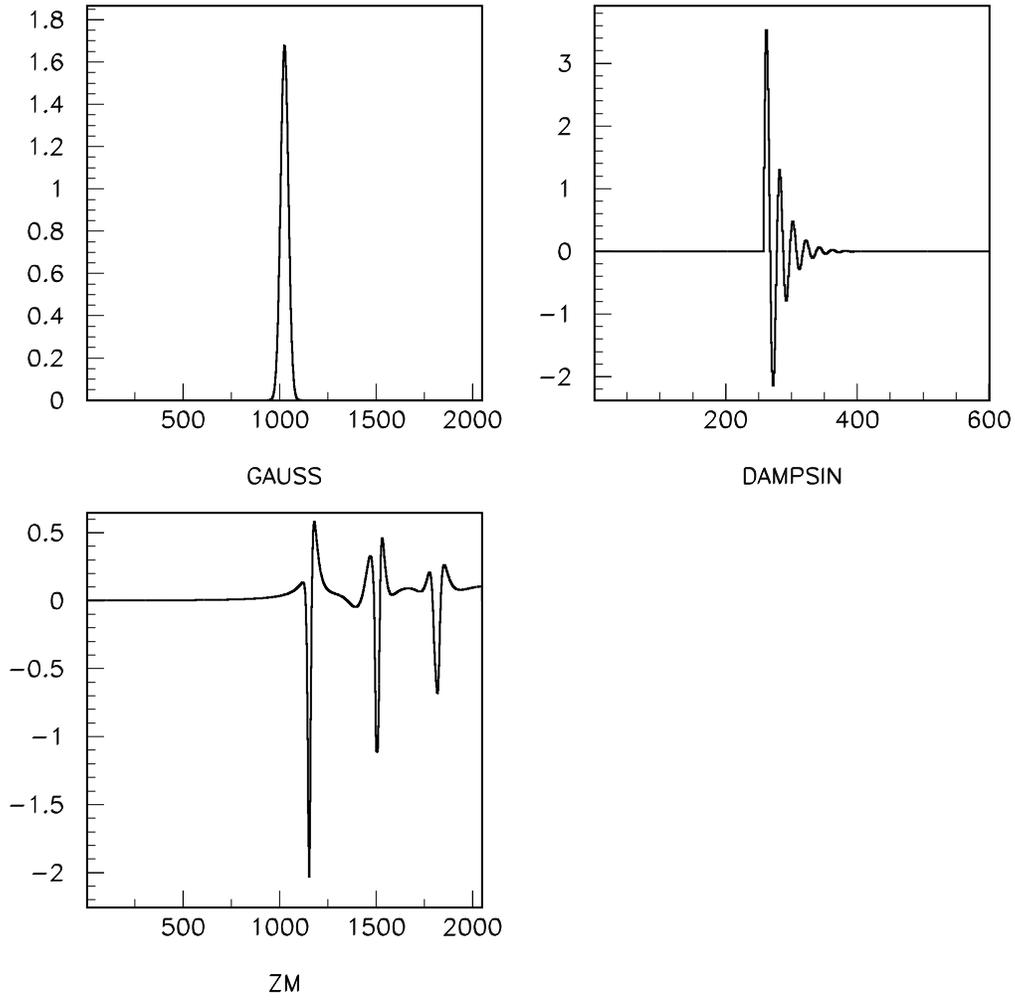,height=15cm}} 
\caption{The signals used for the ROC. Upper-left : Gaussian pulse with
half width 1ms. Upper-right : damped sine with frequency 1 kHz and damping
time 1 ms. Lower : signal emitted by core collapse as simulated by Zwerger
and M\"uller. In the plots, each signal would have an optimal SNR $\rho_0 =10$
if added to a white noise with unity RMS.}

\end{figure}
%----------------------------------------------------------------------------
%---------------------------------FIG 3--------------------------------------
\begin{figure}
\centerline{\epsfig{file=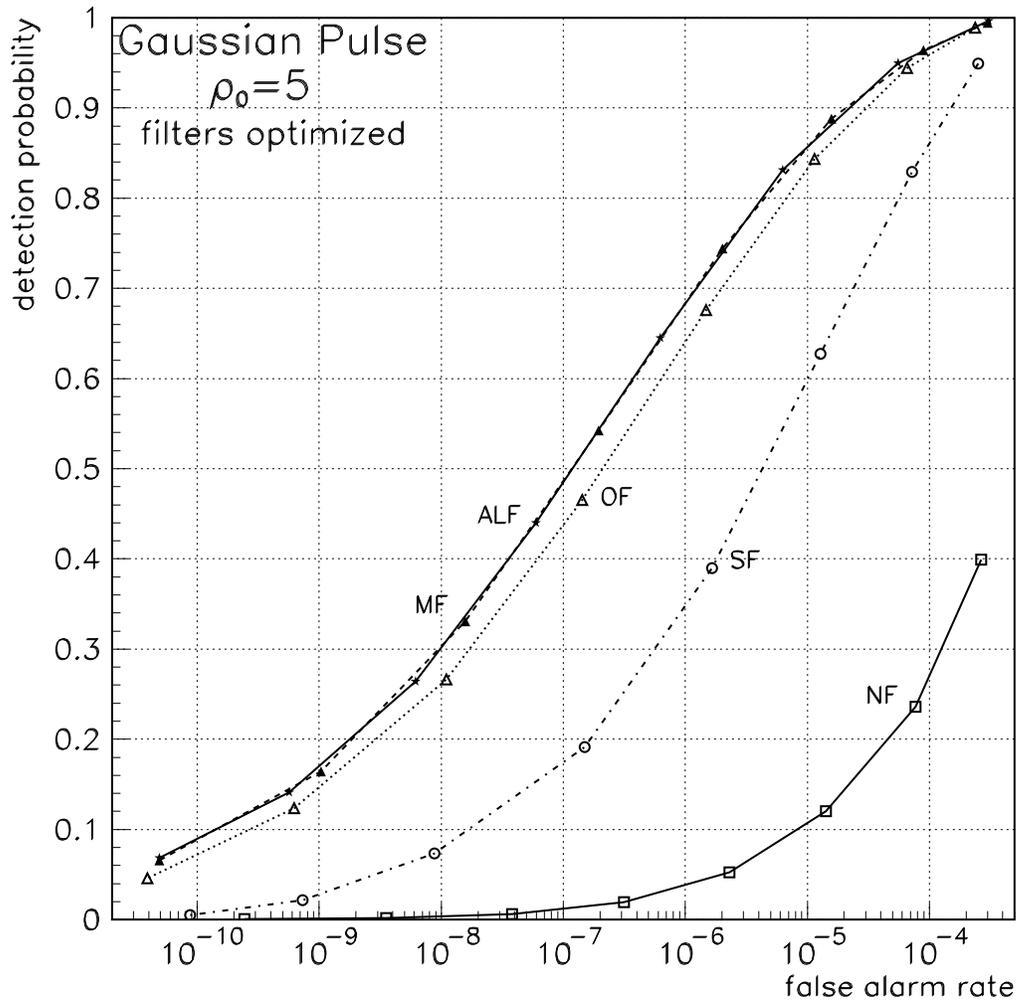,height=15cm}} 
\caption{ROC for optimized filters. The signal is a Gaussian pulse of half-width
1 ms with optimal SNR $\rho_0 = 5$. Black star ($\star$): ALF, white triangle ($\triangle$): OF, white
circle ($\circ$): SF, black triangle ($\blacktriangle$): MF and white square ($\Box$): NF. The false alarm rate
is a false alarm probability {\it per bin}, as in all other ROC.}

\end{figure}
%----------------------------------------------------------------------------
%---------------------------------FIG 4--------------------------------------
\begin{figure}
\centerline{\epsfig{file=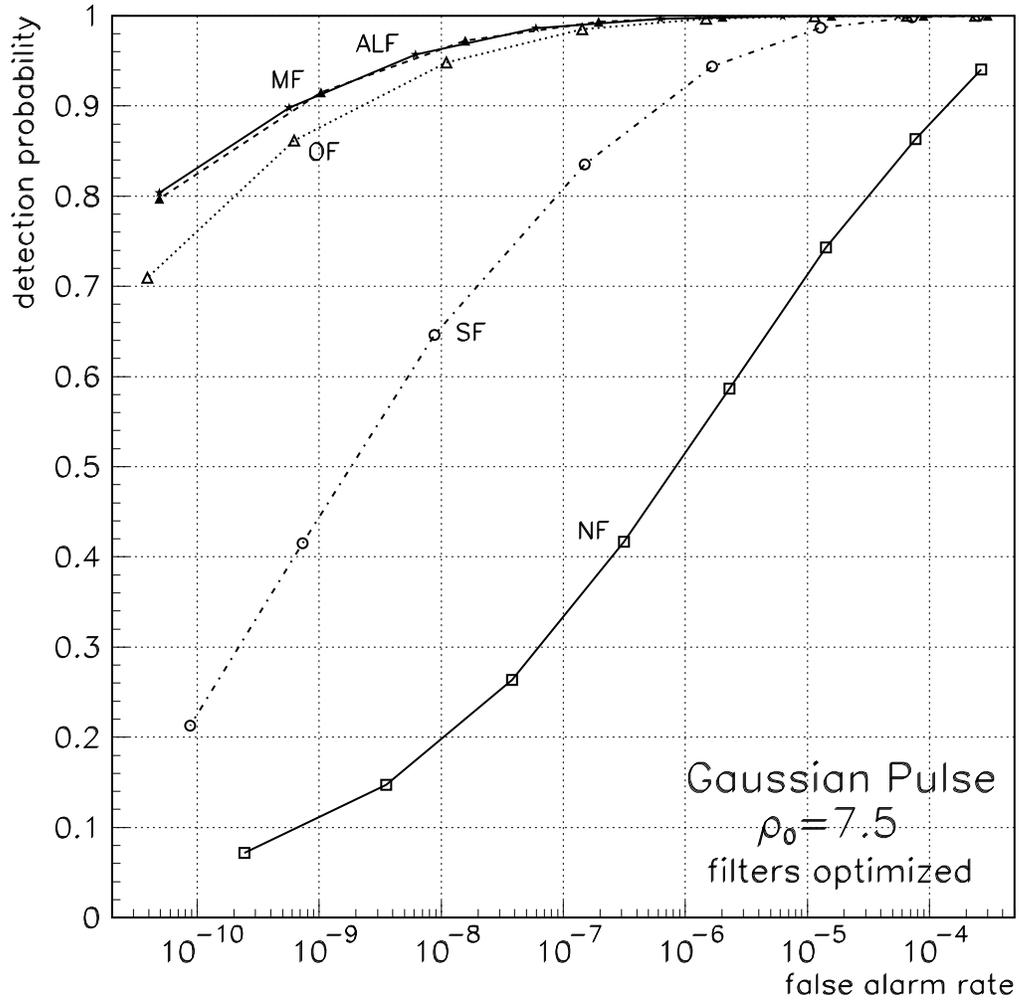,height=15cm}} 
\caption{ROC for optimized filters. The signal is a Gaussian pulse of half-width
1 ms with optimal SNR $\rho_0 = 7.5$. The symbols are $\star$ (ALF), $\triangle$ (OF), $\circ$ (SF), 
$\blacktriangle$ (MF) and $\Box$ (NF).}
\end{figure}
%----------------------------------------------------------------------------
%---------------------------------FIG 5--------------------------------------
\begin{figure}
\centerline{\epsfig{file=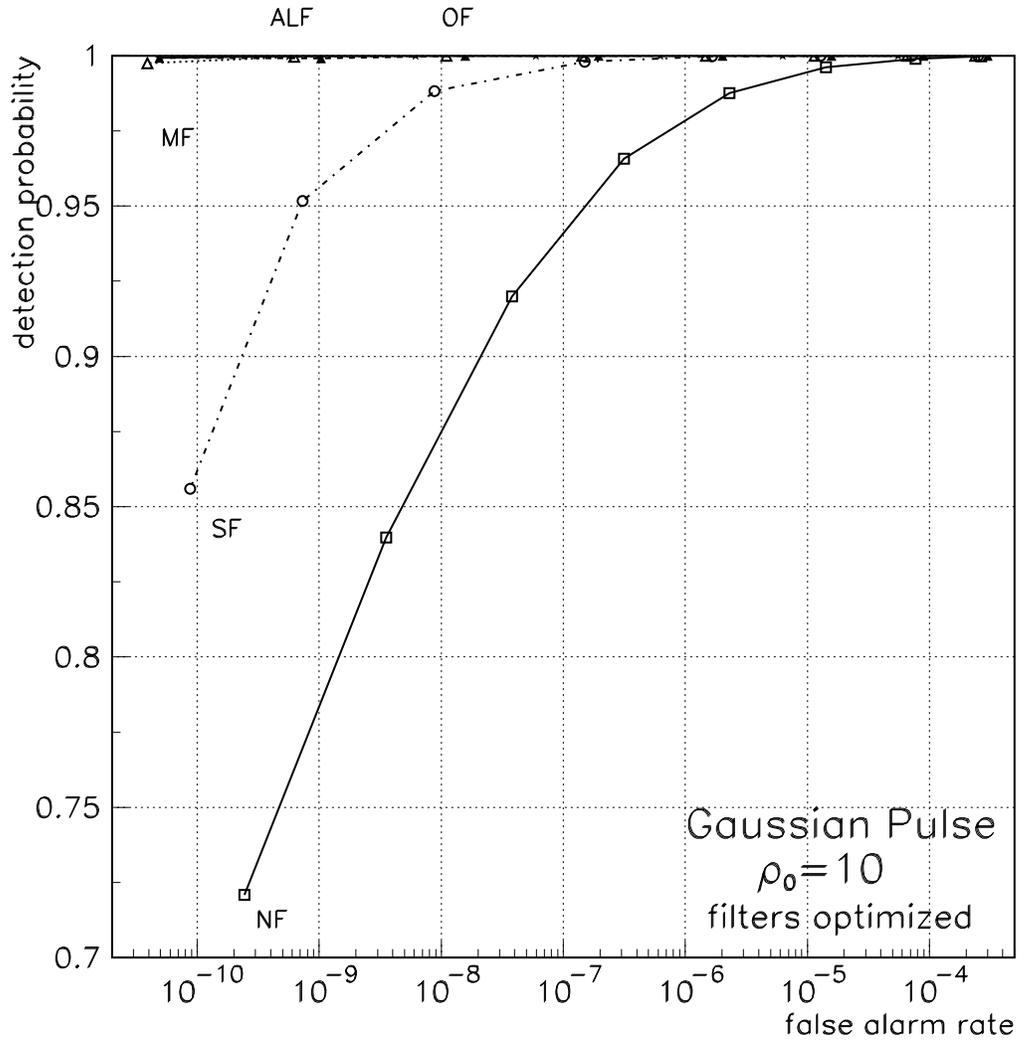,height=15cm}} 
\caption{ROC for optimized filters. The signal is a Gaussian pulse of half-width
1 ms with optimal SNR $\rho_0 = 10$. The symbols are $\star$ (ALF), $\triangle$ (OF), $\circ$ (SF), 
$\blacktriangle$ (MF) and $\Box$ (NF). For such a signal amplitude,
ALF, OF and NF have efficiencies very close to 1, even for very small false alarm rates.}
\end{figure}
%----------------------------------------------------------------------------
%---------------------------------FIG 6--------------------------------------
\begin{figure}
\centerline{\epsfig{file=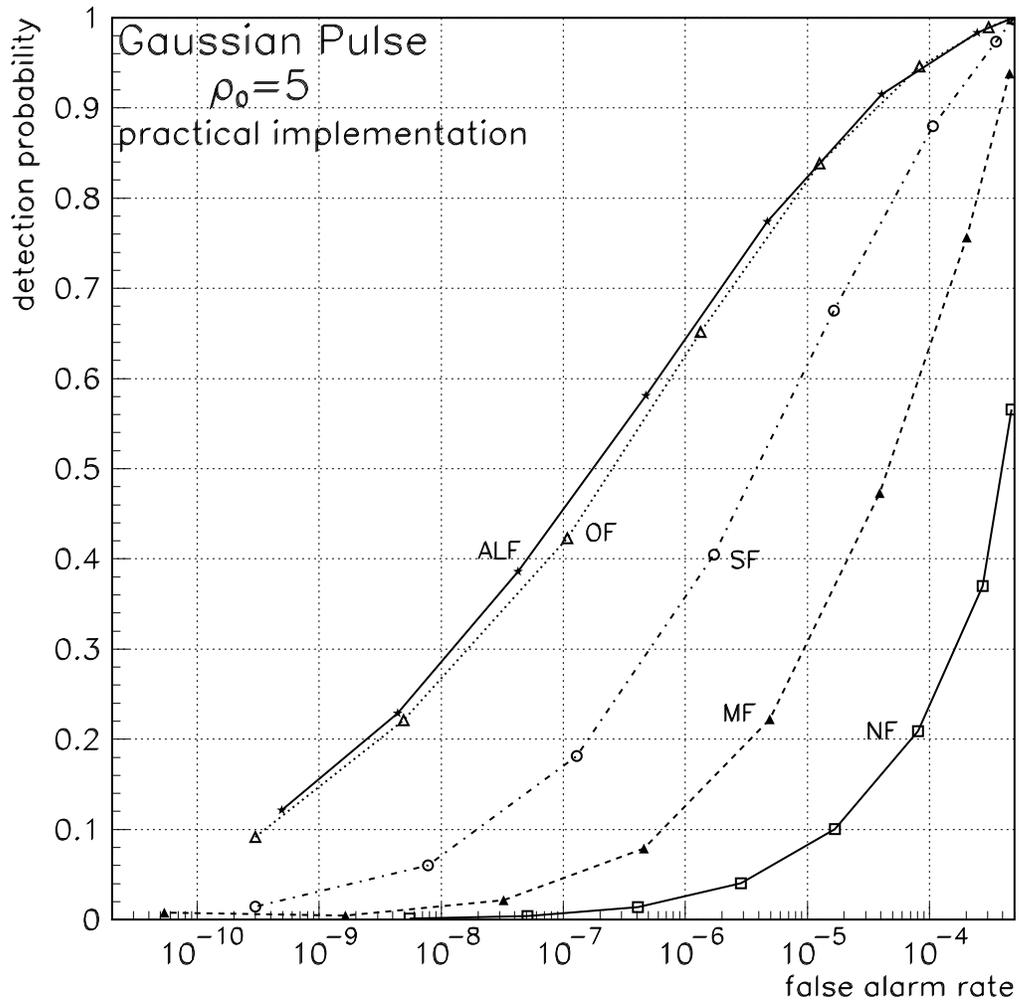,height=15cm}} 
\caption{ROC for filters in a realistic implementation. The signal is a Gaussian pulse of half-width
1 ms with optimal SNR $\rho_0 = 5$. The symbols are $\star$ (ALF), $\triangle$ (OF), $\circ$ (SF), 
$\blacktriangle$ (MF) and $\Box$ (NF).}
\end{figure}
%----------------------------------------------------------------------------
%---------------------------------FIG 7--------------------------------------
\begin{figure}
\centerline{\epsfig{file=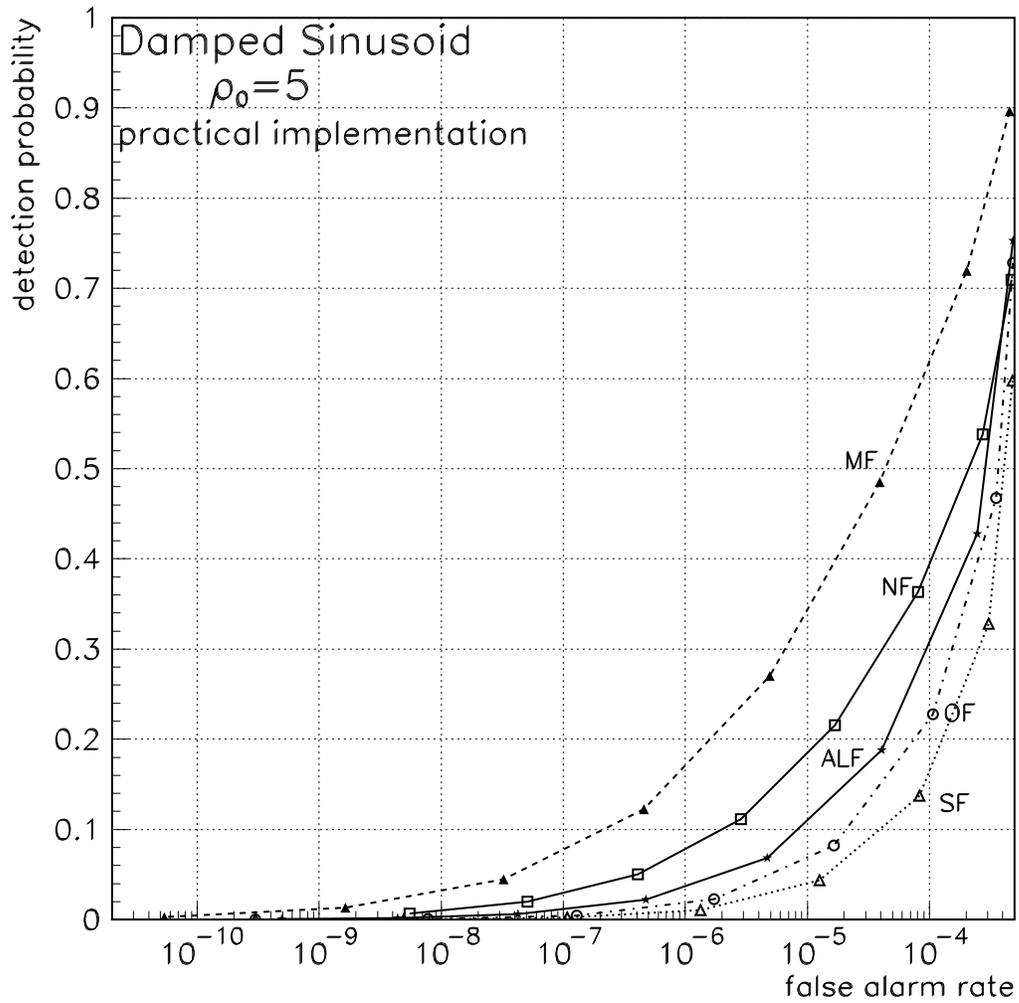,height=15cm}} 
\caption{ROC for filters in a realistic implementation. The signal is a damped sine of frequency
1 kHz and damping time 1ms with optimal SNR $\rho_0 = 5$. 
The symbols are $\star$ (ALF), $\triangle$ (OF), $\circ$ (SF), 
$\blacktriangle$ (MF) and $\Box$ (NF).}
\end{figure}
%----------------------------------------------------------------------------
%---------------------------------FIG 8--------------------------------------
\begin{figure}
\centerline{\epsfig{file=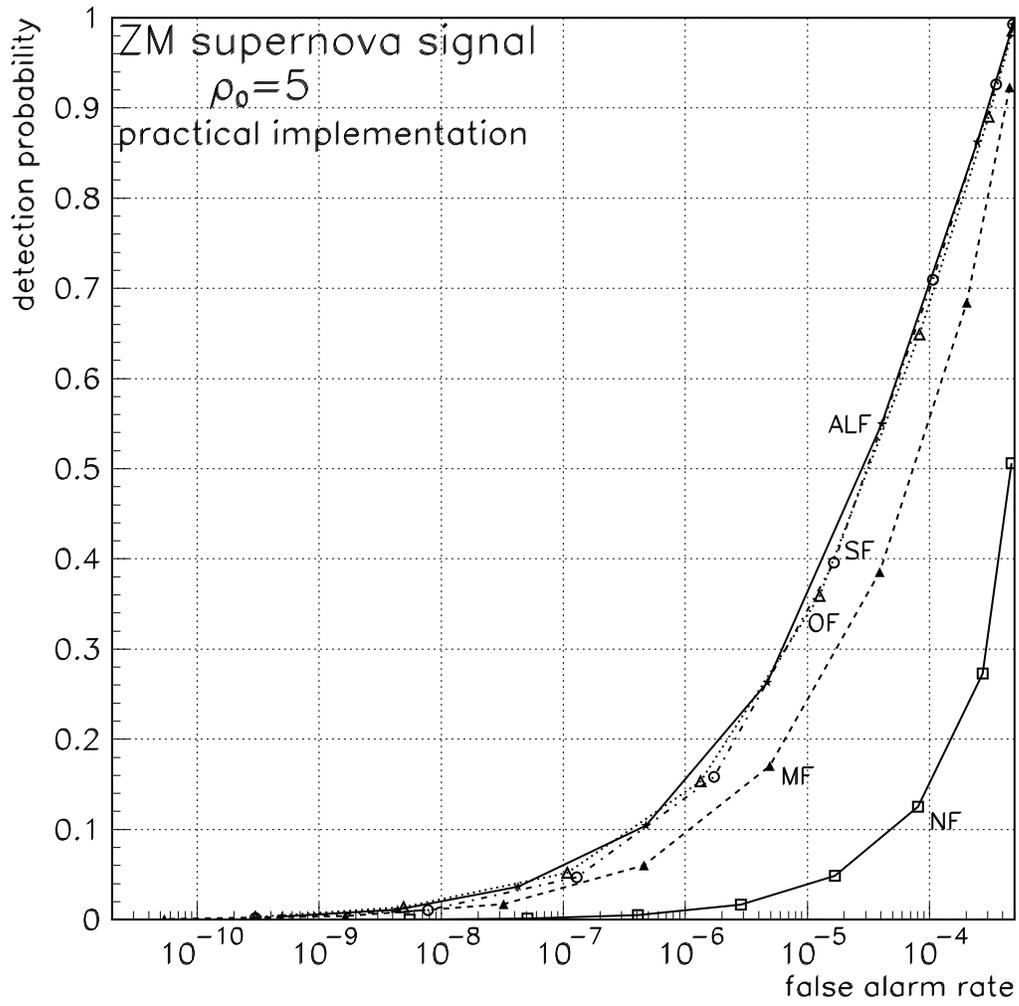,height=15cm}} 
\caption{ROC for filters in a realistic implementation. The signal is a supernova simulated signal
with optimal SNR $\rho_0 = 5$. 
The symbols are $\star$ (ALF), $\triangle$ (OF), $\circ$ (SF), 
$\blacktriangle$ (MF) and $\Box$ (NF).}
\end{figure}
%----------------------------------------------------------------------------
%---------------------------------FIG 9 --------------------------------------
\begin{figure}
\centerline{\epsfig{file=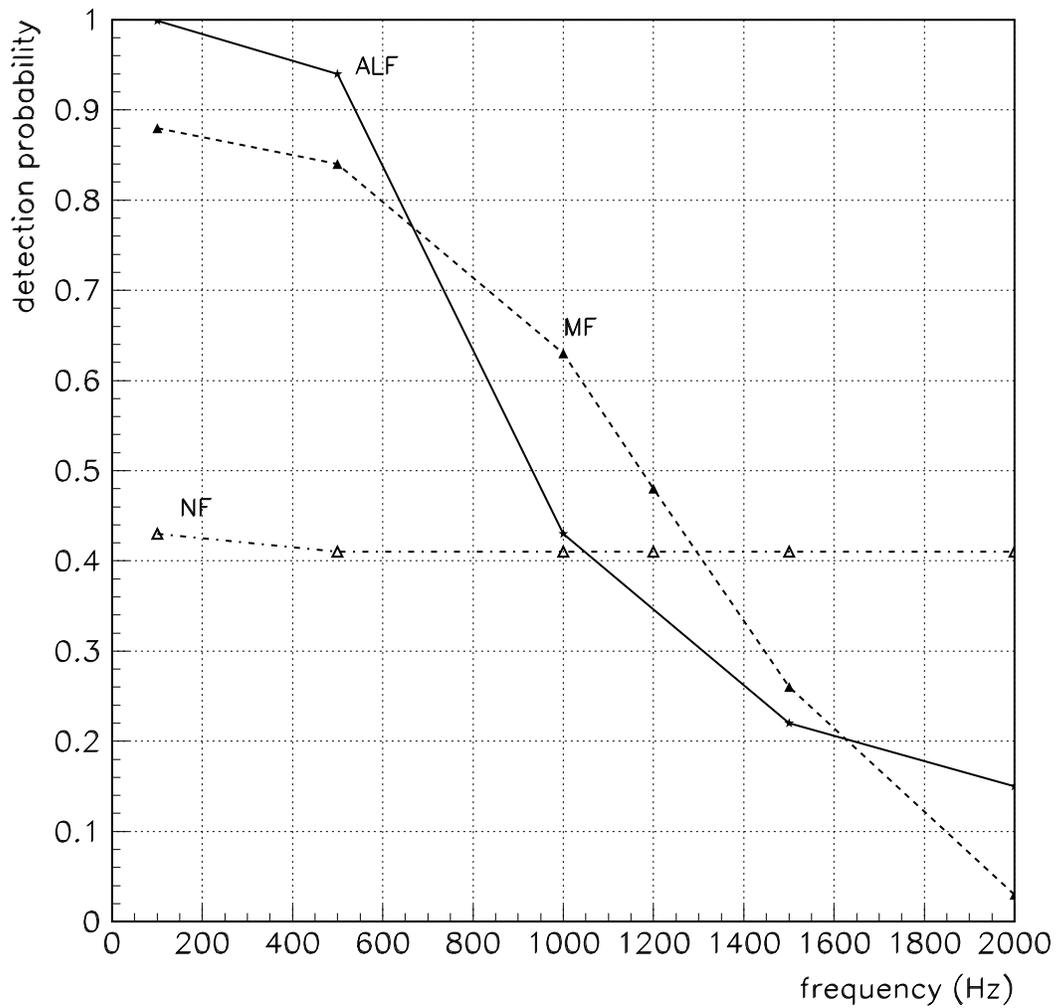,height=15cm}} 
\caption{Detection efficiency of MF, NF and ALF in their practical implementation for a damped sine signal
of damping time $\tau=100$ ms and varying frequency. The signals optimal SNR is $\rho_0 = 5$
and the filter thresholds used here correspond
to a common false alarm rate of about $5\times 10^{-4}$. We note the robustness of NF while ALF and NF
efficiencies decrease with increasing frequency.}
\end{figure}

%----------------------------------------------------------------------------
%---------------------------------FIG 10--------------------------------------
\begin{figure}
\centerline{\epsfig{file=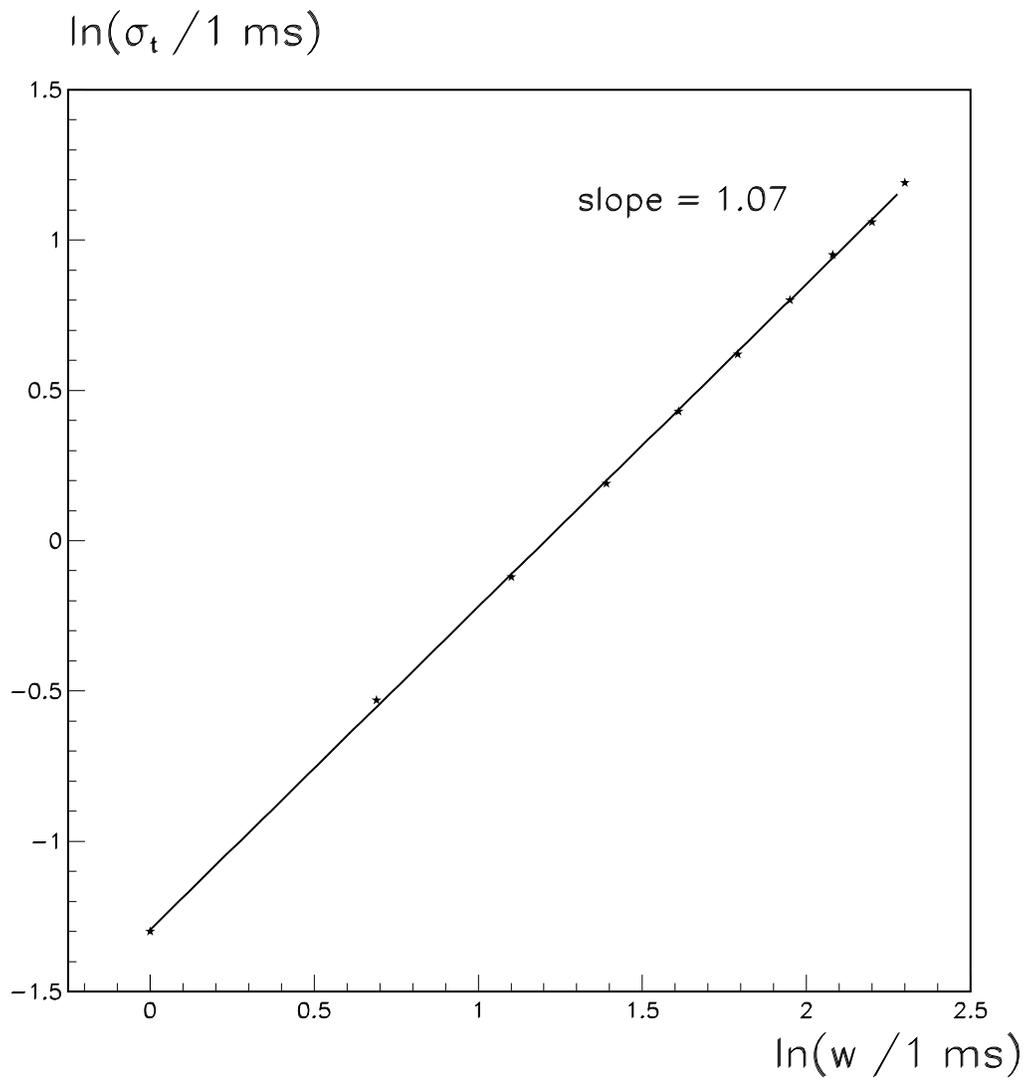,height=15cm}} 
\caption{The NF statistical error for time of arrival reconstruction as a function
of the signal duration in log-log scales. The slope is about 1.08, larger than for the optimal
filter as well as for all the other filters.}
\end{figure}

%----------------------------------------------------------------------------
%---------------------------------FIG 11--------------------------------------
\begin{figure}
\centerline{\epsfig{file=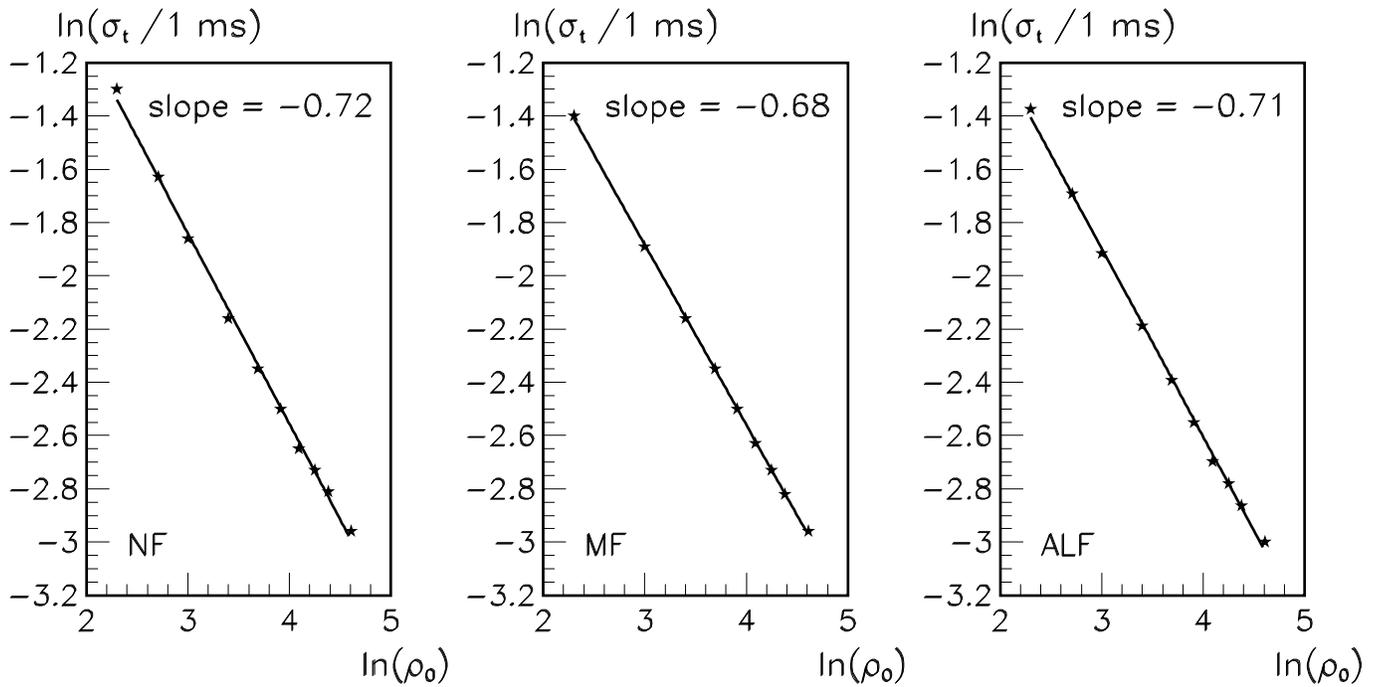,height=10cm}} 
\caption{The statistical errors for time of arrival reconstruction as a function
of the signal amplitude $\rho_0$ in log-log scales for NF, MF and ALF.
In the case of NF, the slope is about -0.72, worse than for the optimal
filter, while it is about -0.68 for MF (a little worse than NF) and about -0.71 for ALF 
(similar to NF).}
\end{figure}

%----------------------------------------------------------------------------
%---------------------------------FIG 12--------------------------------------
\begin{figure}
\centerline{\epsfig{file=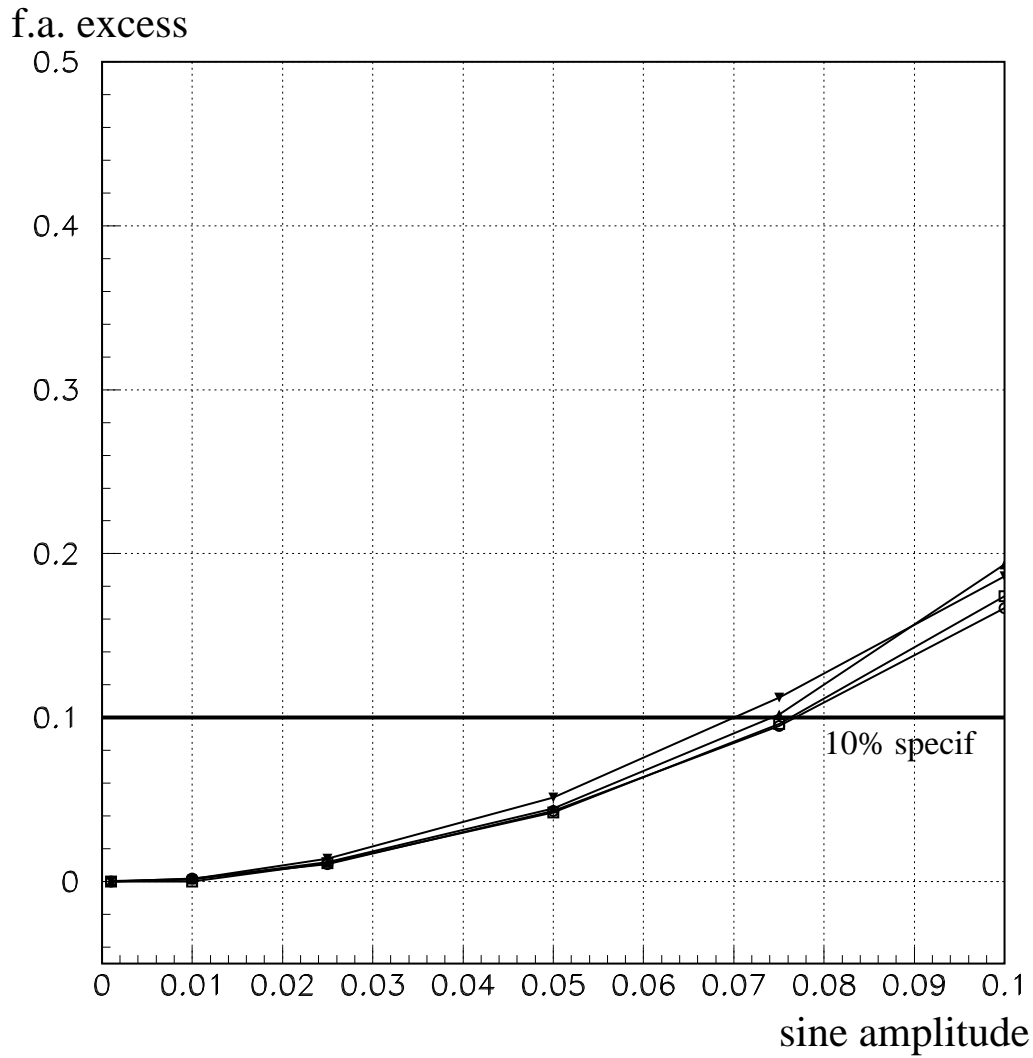,height=15cm}} 
\caption{The relative false alarm excess for the Norm Filter (implemented with $N=50$) for lines of different
frequencies as a function of the line amplitude measured in $\sigma$ units (the Gaussian white noise
RMS). The different frequencies are 0.6 Hz ($\blacktriangle$), 100 Hz ($\blacktriangledown$), 200 Hz
($\circ$) and 400 Hz ($\Box$). We note that the false alarm excess is grossly independant of the line frequency.}
\end{figure}
%----------------------------------------------------------------------------
%---------------------------------FIG 13--------------------------------------
\begin{figure}
\centerline{\epsfig{file=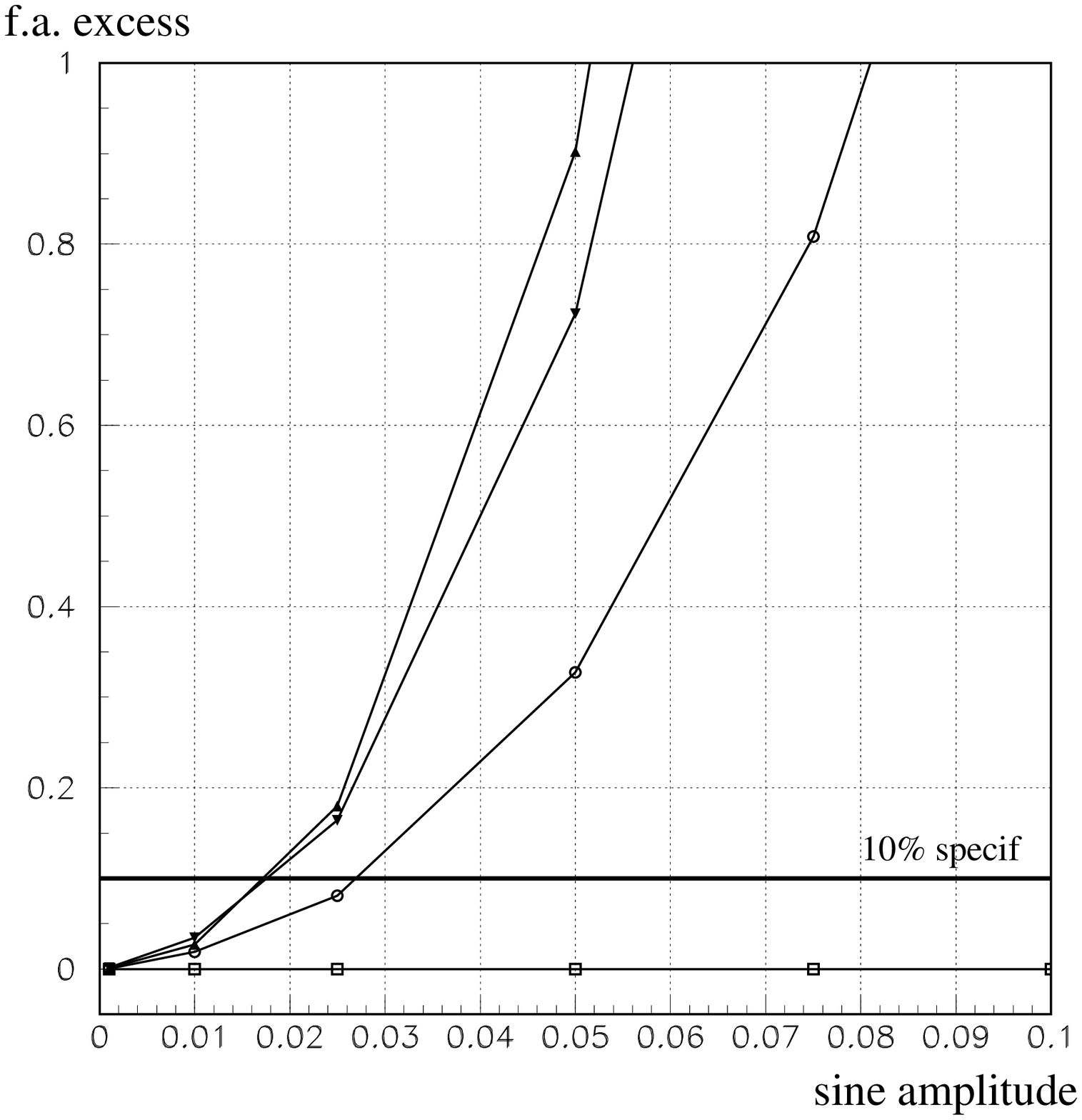,height=15cm}} 
\caption{The relative false alarm excess for the Mean Filter (implemented with $N=50$) for lines of different
frequencies as a function of the line amplitude measured in $\sigma$ units (the Gaussian white noise
RMS). The different frequencies are 0.6 Hz ($\blacktriangle$), 
100 Hz ($\blacktriangledown$), 200 Hz ($\circ$) and 400 Hz ($\Box$). We note that the false alarms excess
decreases with increasing frequencies, until it completely vanishes for frequencies above
the cut-off frequency corresponding to the window size $N$. Here $N=50$ corresponds to a time szie of 2.5 ms
so to a cut-off frequency of 400 Hz.}
\end{figure}
%----------------------------------------------------------------------------
%---------------------------------FIG 14--------------------------------------
\begin{figure}
\centerline{\epsfig{file=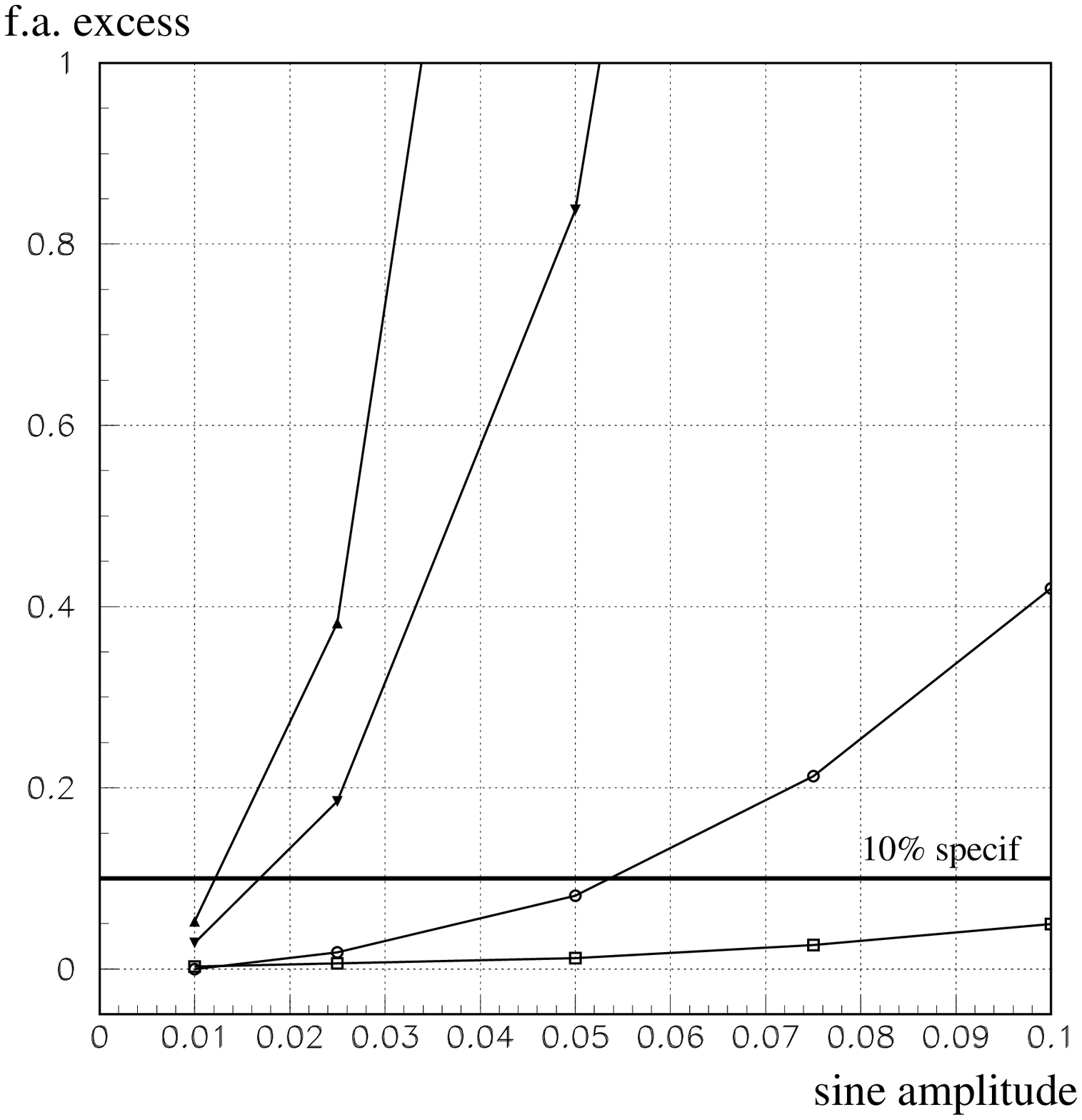,height=15cm}} 
\caption{The relative false alarm excess for ALF (implemented with $N=170$ that is matched to 2.5 ms
signals as for MF and NF) for lines of different
frequencies as a function of the line amplitude measured in $\sigma$ units (the Gaussian white noise
RMS). The different frequencies are 0.6 Hz ($\blacktriangle$), 
100 Hz ($\blacktriangledown$), 200 Hz ($\circ$) and 400 Hz ($\Box$). We note that the false alarms excess
decreases with increasing frequencies.}
\end{figure}

%============================================================================

\begin{references}

%-------------------------------------------------------------------
\bibitem{ligo} A. Abramovici, W.E. Althouse, R.W.P. Drever, Y. G\"ursel,
S. Kawamura, F.J. Raab, D. Shoemaker, L. Sievers, R.E. Spero, K.S. Thorne,
R.E. Vogt, R. Weiss, S.E. Whitcomb and M.E. Zucker, 
Science {\bf 256,} 325 (1992).
%---------------------------------------------------------------------------
\bibitem{virgo} B. Caron {\it et al.} 
``The VIRGO Interferometer for Gravitational Wave Detection'', 
Nucl. Phys. B (Proc. Sup.) {\bf 54B,} 167 (1997).
%-------------------------------------------------------------------
\bibitem{geo} K. Danzmann {\it et al.}, ``GEO 600, a 600 m laser interferometric 
gravitational wave antenna'' in ``Gravitational  wave experiments'',
edited by E. Coccia, G. Pizzella and F. Ronga 
(World Scientific, Singapore, 1995).
%---------------------------------------------------------------------------
\bibitem{tama} Masaki Ando et al., Phys.Rev.Lett. {\bf 86} 3950 (2001).
%---------------------------------------------------------------------------
%\bibitem{thorne87} K.S. Thorne, ``Gravitational radiation`` in ``300 
%years of gravitation'', 
%edited by S.W. Hawking 
%and W. Israel (Cambridge University Press, Cambridge, 1987).
%\bibitem{bona} S. Bonazzola and J.-A. Marck, Annu.Rev.Nucl.Part.Sci. 
%{\bf 45,} 655 (1994).
\bibitem{schutz} B.F. Schutz, ``Data processing, analysis and storage 
for interferometric antennas'', 
in ``The detection of gravitational waves'', edited by D.G. Blair 
(Cambridge University Press, Cambridge, 1991).
\bibitem{monchmeyer} R. M\"{o}nchmeyer, G. Sch\"{a}fer, 
E. M\"uller and R.E. Kates, Astron. Astrophys. {\bf 246,} 417 (1991).
\bibitem{bona1} S. Bonazzola and J.-A. Marck, 
Astron. Astrophys. {\bf 267,} 623 (1993).
\bibitem{zwerger} T. Zwerger and E. M\"uller, 
Astron. Astrophys. {\bf 320,} 209 (1997).
\bibitem{rampp} M. Rampp, E. M\"uller and M. Ruffert, 
Astron. Astrophys. {\bf 332,} 969 (1998).
\bibitem{relat} H. Dimmelmeier, J. A. Font and E. M\"uller,
Astrophys. J. {\bf 560,} L163 (2001).
\bibitem{relat2} H. Dimmelmeier, J. A. Font and E. M\"uller,
``Gravitational waves from relativistic rotational core collapse
in axisymmetry'', 4th Edoardo Amaldi Conference on Gravitational Waves (Perth July8-14 2001), to appear in Class. Quantum. Grav. 
\bibitem{relat3} H. Dimmelmeier, J. A. Font and E. M\"uller,
``Relativistic simulations of rotational core collapse. II. 
Collapse dynamics and gravitational radiation'', astro-ph/0204289 (2002).
%\bibitem{Cordes} J.M. Cordes, R.W. Romani and S.C. Lundgren, 
%Nature {\bf 362,} 133 (1993).
%\bibitem{Nazin} S.N. Nazin and K.A. Postnov, Astron. Astrophys. 
%{\bf 317,} L79 (1997).
\bibitem{stark} R.F. Stark and T. Piran, Phys. Rev. Lett. {\bf 55,} 891 (1985).
                
\bibitem{arnaud} N. Arnaud, F. Cavalier, M. Davier and P. Hello, 
Phys. Rev. D {\bf 59,} 082002 (1999).

\bibitem{ooh} K. Oohara and T. Nakamura, ``Coalescences of binary neutron 
stars'' 
in ``Relativistic gravitation and gravitational waves'', edited by
J.-A. Marck and J.-P. Lasota (Cambridge University Press, Cambridge, 1997).
\bibitem{rasio} F.A. Rasio and S.L. Shapiro, Class.Quant.Grav. {\bf 16,} R1 (1999).
\bibitem{ruf} M. Ruffert and H.-Th. Janka, 
Astron. Astrophys. {\bf 338,} 535 (1998).
\bibitem{janka} H.-T. Janka, T. Eberl, M. Ruffert and C.L. Fryer, 
Astrophys. J. {\bf 527,} L39 (1999).
\bibitem{jr} H.-T. Janka and M. Ruffert, ``Detectable signals from mergers
of compact stars'', in the proceedings of the conference on Stellar collisons,
mergers and their consequences, ASP Conference series, ed. M. Shara, in press.
\bibitem{jr2} H.-T. Janka and M. Ruffert, `` Coalescing neutron stars - 
a step towards physical models III. Improved numerics and different
 neutron star masses and spins'', astro-ph/0106229, accepted for publication
in Astron. Astrophys. (2001).



%\bibitem{FMR} L.S. Finn, S.D. Mohanty and J.D. Romano,
%Phys. Rev. D {\bf 60,} 121101 (1999).



\bibitem{bbh1} M. Alcubierre, W. Benger, B. Bruegmann, G. Lanfermann, L. Nerger, 
E. Seidel and  R. Takahashi, Phys.Rev.Lett. {\bf 87,} 271103 (2001).
\bibitem{bbh2} M. Shibata and K. Uryu, Prog.Theor.Phys. {\bf 107,}  265 (2002).
\bibitem{bbh3} J. Baker, M. Campanelli and C. Lousto, 
Phys.Rev. D {\bf 65,} 044001 (2002).
\bibitem{bbh4} J. Baker, M. Campanelli, C. Lousto and R. Takahashi,
Phys.Rev. D {\bf 65,} 124012 (2002).
\bibitem{bbh5} S. Husa, Y. Zlochower, R. Gomez and J. Winicour,
Phys.Rev. D {\bf 65,} 084034 (2002).
%\bibitem{det} S. Detweiler, Astrophys. J. {\bf 239,} 292 (1980).
%\bibitem{ech} F. Echeverria, Phys. Rev. D {\bf 40,} 3194 (1989).
%\bibitem{san} S.V. Dhurandhar and M. Tinto, 
%Mon. Not. R. astr. Soc. {\bf 234,} 663 (1988).


\bibitem{cusps} T. Damour and A. Vilenkin, Phys.Rev. D  {\bf 64} 064008 (2001). 

\bibitem{flana} E.E. Flanagan and S.A Hughes, 
Phys. Rev. D {\bf 57,} 4535 (1998).
\bibitem{powermonit} W. G. Anderson, P.R. Brady, J.D.E Creighton and 
E.E. Flanagan, Int.J.Mod.Phys. D {\bf 9,}  303 (2000).
\bibitem{power2} W. G. Anderson, P.R. Brady, J.D.E Creighton and 
E.E. Flanagan, Phys. Rev. D {\bf 63,} 042003  (2001).
\bibitem{vicere} A. Vicer\'e, Phys. Rev. D 66, 062002 (2002)

\bibitem{bala} W. G. Anderson and R. Balasubramanian, 
Phys. Rev. D {\bf 60,} 102001 (1999).
\bibitem{mohanty} S. D. Mohanty Phys. Rev. D {\bf 61,} 122002 (2000). 
\bibitem{fabbroni} L. Fabbroni and M. Vannucci, ``Wavelet tests for the detection
of transients'', VIRGO note VIR-NOT-FIR 1390 151 (2000).
\bibitem{sylvestre} J. Sylvestre, ``Time-frequency detection algorithm for 
gravitational wave bursts'', gr-qc/0210043.

\bibitem{moriond}N. Arnaud, F. Cavalier, M. Davier, P. Hello and T. Pradier,
''Triggers for the detection of gravitational wave bursts'', to appear in the 
proceedings of the XXXIVth Rencontres de Moriond on "Gravitational Waves and 
Experimental Gravity" (les Arcs, Jan.99), gr-qc/9903035.
\bibitem{pra} T. Pradier, N. Arnaud, M.-A. Bizouard, F. Cavalier, M. Davier 
and P. Hello, Int.J.Mod.Phys. D {\bf 9,} 309 (2000).
\bibitem{prados} T. Pradier, N. Arnaud, M.-A. Bizouard, F. Cavalier, M. Davier 
and P. Hello, Phys. Rev. D {\bf 63,} 042002 (2001).

\bibitem{papou} A. Papoulis, ``Signal Analysis'', p.343, McGraw-Hill,New York, 1977.

\bibitem{pradthese} T. Pradier, PhD thesis (LAL 01-15, Orsay, 2001).

\bibitem{virgosens} http://www.virgo.infn.it/senscurve/

\bibitem{cuo}E. Cuoco, G. Calamai, L. Fabbroni, G. Losurdo, M. Mazzoni, R. Stanga, 
F. Vetrano, Class.Quant.Grav. {\bf 18,} 1727 (2001).
\bibitem{cuo2} E. Cuoco, G. Losurdo, G. Calamai, 
L. Fabbroni, M. Mazzoni, R. Stanga, G. Guidi, 
F. Vetrano, Phys.Rev. D {\bf 64,} 122002 (2001).
\bibitem{spiegel} M.R. Spiegel, ``Probabilit\'es et statistiques'',
McGraw-Hill, Paris, 1981.
\bibitem{webSN} http://www.mpa-garching.mpg.de/$\sim$ewald/GRAV/grav.html

\bibitem{neutrinos} N. Arnaud, M. Barsuglia, M.-A. Bizouard, F. Cavalier, M. Davier,
P. Hello and T. Pradier,  Phys. Rev. D {\bf 65,} 033010 (2002).

\bibitem{networknous} N. Arnaud, M. Barsuglia, M.-A. Bizouard, P. Canitrot, F. Cavalier, M. Davier,
P. Hello and T. Pradier,  Phys. Rev. D {\bf 65,} 042004 (2002).


\end{references}
\end{document}